\documentclass[aps,showpacs,showkeys,preprintnumbers]{revtex4}
\usepackage{graphicx}
\usepackage{color}
\usepackage{amsmath}
\usepackage{amssymb}
\usepackage{multirow}
\usepackage{titlesec}
\makeatletter
\renewcommand\paragraph{\@startsection{paragraph}{4}{\z@}%
            {-2.5ex\@plus -1ex \@minus -.25ex}%
            {1.25ex \@plus .25ex}%
            {\normalfont\normalsize\bfseries}}
\makeatother
\def \be  {\begin{equation}}
\def \ee  {\end{equation}}
\def \ee  {\end{equation}}
\def \bea {\begin{eqnarray}}
\def \eea {\end{eqnarray}}

\newcommand{\Kslxi}{$K^{0}_{s}$, $\Lambda$, and $\Xi^{-}$}

\begin{document}

\title{Transverse momentum spectra of strange hadrons within extensive and nonextensive statistics}

\author{Hayam Yassin}
\email{hiam_hussien@women.asu.edu.eg}
\affiliation{Physics Department, Faculty of Women for Arts, Science and Education, Ain Shams University, 11577 Cairo, Egypt}

\author{Eman R. Abo Elyazeed}
\email{eman.reda@women.asu.edu.eg}
\affiliation{Physics Department, Faculty of Women for Arts, Science and Education, Ain Shams University, 11577 Cairo, Egypt}

\author{Abdel Nasser  Tawfik}
\email{atawfik@nu.edu.eg}
\affiliation{Nile University - Egyptian Center for Theoretical Physics (ECTP), Juhayna Square off 26th-July-Corridor, 12588 Giza, Egypt}

\begin{abstract}

Using generic (non)extensive statistics, in which the underlying system autonomously manifests its extensive and nonextensive statistical nature, we extract various fit parameters from the \textit{CMS}  experiment and compare these to the corresponding results obtained from Tsallis and Boltzmann statistics. The present study is designed to indicate the possible variations between the three types of statistical approaches and characterizes their dependence on collision energy, multiplicity, and size of the system of interest. We analyze the transverse momentum spectra $p_{\mathrm{T}}$ of the strange hadrons \Kslxi produced in $\textsf{Pb+Pb}$ collisions, at $\sqrt{s_{\mathrm{NN}}}=2.76~$TeV, in $\textsf{p+Pb}$ collisions, at $\sqrt{s_{\mathrm{NN}}}=5.02~$TeV, and in $\textsf{p+p}$ collisions, at $\sqrt{s_{\mathrm{NN}}}=7~$TeV. From the comparison of the resulting fit parameters; temperature $T$, volume $V$, and nonextensvie parameter $d$, with calculations based on Tsallis and Boltzmann statistics, remarkable differences between the three types of statistics are determined besides a strong dependence on size and type of the colliding system. We conclude that the produced particles with large masses and large strange quantum numbers likely freeze out earlier than the ones with smaller masses and less strange quantum numbers. This conclusion seems not depending on the type of the particle or the collision but apparently manifesting transitions from chemical (larger temperature) to the kinetic freezeouts (lower temperature). For the first university (equivalent) class $c\sim1$, the decrease in the second one, $d$, with increasing energy and collision centrality highlights that the system departs from nonextensivity (non-equilibrium) and apparently approaches extensivity (equilibrium) indicating that the Boltzmann statistics becomes the proper statistical approach in describing that system. Last but not least, we present analytical expressions for the energy dependence of the various fit parameters.

\end{abstract}


\date{\today}

\maketitle


\section{Introduction}
\label{intro}

The high-energy experiments are designed to study the strongly interacting matter, at high temperatures and densities \cite{Khuntia:2017ite}. The deconfinement of colliding hadrons into quark-gluon plasma (QGP), which then rapidly expands and cools down \cite{Ahmad:2013una}, is conjectured to be created at such extreme collisions  \cite{Bjorken:1982qr,Ullrich:2013qwa,Gyulassy:2004zy,Tawfik:2014eba}. There are different signatures for the formation of QGP. The enhancement of the strangeness number \cite{Rafelski:1994gi,ALICE:2017jyt} and of the transverse momentum $p_{\mathrm{T}}$ spectra of strange particles \cite{Song:2010mg} are well-known examples. The importance of the latter is the ability to determine the freezeout parameters; the temperature and the chemical potential in different statistical approaches \cite{Adams:2003xp,Becattini:2004td,Mueller:1993rr,Mueller:1994jq,Mueller:1994gb,Schnedermann:1993ws,Tawfik:2017bul}. These statistical models are able to describe the experimental results, at a wide range of energies \cite{Tawfik:2014eba}.

So far, various phenomena have been successfully described by extensive (Boltzmann) \cite{Hagedorn:1965st,Tawfik:2014eba,Tawfik:2010aq}, nonextensive Tsallis \cite{Tawfik:2013msa,Tawfik:2010pt,Tawfik:2010aq}, and generic (non)extensive statistical approach \cite{Beck:2000nz,Bediaga:1999hv,Tawfik:2016pwz,Tawfik:2016jol,Tawfik:2017bsy,Tawfik:2018ahq}. Focusing the discussion on the transverse momentum spectra $p_{\mathrm{T}}$, there are different Tsallis approaches  \cite{Tsallis:1987eu,Biro:2008hz,Bhattacharyya:2015hya,Zheng:2015gaa,Tang:2008ud,De:2014dna} which have been utilized in characterizing the heavy-ion collisions \cite{Adare:2011vy,Khachatryan:2011tm,Gao:2015qsq}. Also, the $p_{\mathrm{T}}$ spectra of well-identified particles in $\textsf{p+p}$ collisions, at the Relativistic Heavy Ion Collider (RHIC) and the Large Hadron Collider (LHC) energies have been excellently fitted to Tsallis statistics \cite{Khandai:2013gva,Adare:2010fe,Sett:2014csa}. Furthermore, the Tsallis approaches are used in analyzing $p_{\mathrm{T}}$ of charged and strange hadrons \cite{Khandai:2013fwa,Saraswat:2017gqt,Cleymans:2011in,Tang:2008ud,De:2014dna,Sett:2015lja}. The generic (non)extensive approach, which was introduced in ref. \cite{Tawfik:2016pwz} and utilized in estimating the particle production, at a wide range of energies \cite{Tawfik:2016jol,Tawfik:2017bsy,Tawfik:2018ahq} was utilized to analyze the $p_{\mathrm{T}}$ spectra of charged particles measured in different types of collisions at RHIC and LHC energies \cite{Tawfik:2017bul,Tawfik:2019oct}. This motivates the present work, in which $p_{\mathrm{T}}$ of strange hadrons are fitted to extensive (Boltzmann) and nonextensive approaches [Tsallis and generic (non)extensive], at various energies. The outcomes are  the freezeout parameters; temperature and chemical potential besides the nonextensivity parameter, itself. The latter allows us to estimate the degree of nonextensivity of the underlying system. Accordingly, we are furnished with a trustful tool enabling us to judge about formation of QGP, as this is likely accompanied by critical phenomena and therefore has a nonextensive statistical nature.

The present script focuses on strange particles as they are suitable for the manifestation of the nonextensive statistical nature of the particle production and can compared with the conclusions drawn in refs. \cite{Castorina:2014cia,Tawfik:2016tfe}. They can as well come up with an essential contribution to the conventional dependence of the temperature on the strange quantum numbers, which likely strengthens the signature proposed for the QGP formation.

The present paper is organized as follows. The statistical approaches are given in section \ref{sec:app}. The particle spectra within extensive Boltzmann and generic (non)extensive statistics are presented in section \ref{Boltzmann} and \ref{generic}, respectively. The  results on $p_{\mathrm{T}}$ spectra of strange hadrons measured in different types of collisions are elaborated in section \ref{sec:res}. The dependence of the resulting fit parameters on the collision centralities, the strange particle number, and the type of strange particles are analyzed in section \ref{sec:res}. The dependence of the various fit parameters on the collision energies is discussed in section \ref{res2}. Section \ref{sec:concl} is devoted to the conclusions.

\section{Statistical Approaches}
\label{sec:app}

\subsection{Boltzmann Statistics}
\label{Boltzmann}

We start with the total number of particles in an statistical ensemble as expressed within Boltzmann statistics \cite{Ning:2003},
\begin{eqnarray}
\label{eq1}
N = \frac{g V}{(2\pi)^3} \int_0^\infty \frac{d{p^3}}{\exp\left(\frac{E-\mu}{T_\mathrm{B}}\right)},
\end{eqnarray}
where $T_\mathrm{B}$ is the Boltzmann temperature. The momentum distribution can be deduced as \cite{Ning:2003,Dermer:1984,Meng:2009zzc}
\begin{eqnarray}
\label{eq2}
E \frac{d^3N}{dp^3} = \frac{g V E}{(2\pi)^3} \exp\left(\frac{\mu-E}{T_\mathrm{B}}\right).
\end{eqnarray}
At mid-rapidity $y=0$ and $\mu \approx 0$, the momentum distribution reads
\begin{eqnarray}
\label{eq3}
\left.\frac{1}{2\pi p_{\mathrm{T}}}\frac{d^2N}{dp_{\mathrm{T}}dy}\right|_{y=0}=\frac{g V m_{\mathrm{T}}}{(2\pi)^3} \exp\left(-\frac{m_{\mathrm{T}}}{T_\mathrm{B}}\right).
\end{eqnarray}
where $E=m_{\mathrm{T}} \cosh y$. The transverse mass $m_{\mathrm{T}}$ is  given as $\sqrt{p_T^2 + m^2}$. At mid-rapidity ($y=0$) but $\mu \neq 0$, the momentum distribution can be given as
\begin{eqnarray}
\label{eq4}
\left.\frac{1}{2\pi p_{\mathrm{T}}}\frac{d^2N}{dp_{\mathrm{T}}dy}\right|_{y=0}=\frac{g V m_{\mathrm{T}}}{(2\pi)^3} \exp\left(\mu-\frac{m_{\mathrm{T}}}{T_\mathrm{B}}\right).
\end{eqnarray}

\subsection{Generic (non)extinsive Statistics}
\label{generic}

For a quantum gas with fermion and boson constituents, the partition function within the generic axiomatic-(non)extensive approach reads \cite{Tawfik:2017bsy}
\begin{eqnarray}
\ln\, Z_{\mathtt{FB}}(T) &=& \pm V\, \sum_i^{N_{\mathtt{F|B}}}\, g_i \int_0^{\infty} \frac{d^3\, {\bf p}}{(2\, \pi)^3}\; \ln\left[1\pm\varepsilon_{c,d,r}(x_i)\right], \label{eq5}
\end{eqnarray}
where $\pm$ refer to fermions (subscript $\mathtt{F}$) and bosons (subscript $\mathtt{B}$), respectively. The distribution function $\varepsilon_{c,d,r}(x_i)$ is given as \cite{Thurner1,Hanel_2011}
\begin{equation}
\varepsilon_{c,d,r}(x)=\exp\left[ \frac{-d}{1-c} \left(W_k\left[B\left(1-\frac{x}{r}\right)^{\frac{1}{d}}\right]-W_k[B]\right)\right], \label{eq:epsln}
\end{equation}
where $W_k$ is the Lambert W-function which has real solutions at $k=0$ with $d\geq 0$ and at $k=1$ with $d<0$,
\begin{equation}
B=\frac{(1-c)r}{1-(1-c)r} \exp\left[\frac{(1-c)r}{1-(1-c)r}\right],
\end{equation}
with $r=[1-c+c\,d]^{-1}$ and the universality (equivalent) class $(c,d)$ does not only define the entropy and throughout the extensive and nonextensive statistical nature of the underlying system, but as shown in Eq. (\ref{eq:epsln}) it determines the correspondent distribution function. Thus, the total number of particles can be determined from as
\begin{eqnarray}
N_i &=& \frac{\pm V}{8 \pi^3} g_i \int_0^\infty \frac{ \varepsilon_{c,d,r}(x_i) \; W_0\left[B(1-\frac{x_i}{r})^{\frac{1}{d}}\right]}{(1-c)\left[1\pm\varepsilon_{c,d,r}(x_i)\right] \left(r-x_i\right) \left(1+W_0\left[B(1-\frac{x_i}{r})^{\frac{1}{d}}\right]\right)} d^3\textbf{p}, \label{FDn}
\end{eqnarray}
where $x_i=(\mu-E_i)/T$, $i$ runs over \Kslxi, with $g_{K_s^0}=1$ and $g_\Lambda=g_{\Xi^-}=2$ are degeneracy factors.
The corresponding momentum distribution for strange hadrons is given as
\begin{equation}
\frac{1}{2 \pi}\frac{E d^3N}{d^3p}= \frac{\pm g_i V E_i T}{8 \pi^3} \frac{\varepsilon_{c,d,r}\left(x_i\right) \; W_0\left[B\left(1-\frac{x_i}{r}\right)^{\frac{1}{d}}\right]}{(1-c)\left[1\pm\varepsilon_{c,d,r}\left(x_i\right)\right] \left(r T -\mu+E_i\right) \left(1+W_0\left[B\left(1-\frac{x_i}{r}\right)^{\frac{1}{d}}\right]\right)},
\label{eq:GmuZero}
\end{equation}
where $\varepsilon_{c,d,r}\left(x_i\right)$ is defined in Eq. (\ref{eq:epsln}).

The transverse momentum distribution in terms of rapidity ($y$) and transverse mass (${m_T}_i = \sqrt{p_T^2 + m^2_i}$) can be expressed as
\begin{eqnarray}
&&\frac{1}{2 \pi}\frac{d^2 N}{dy m_T dm_T}= \frac{\pm g_i V T {m_T}_i \cosh y}{8 \pi^3} \times  \\
&&\frac{\varepsilon_{c,d,r}\left(\frac{\mu-{m_T}_i \cosh y}{T}\right) \; W_0\left[B\left(1-\frac{\mu-{m_T}_i \cosh y}{r T}\right)^{\frac{1}{d}}\right]}{(1-c)\left[1\pm\varepsilon_{c,d,r}\left(\frac{\mu-{m_T}_i \cosh y}{T}\right)\right] \left(r T+ {m_T}_i \cosh y-\mu \right) \left(1+W_0\left[B\left(1-\frac{\mu-{m_T}_i \cosh y}{r T}\right)^{\frac{1}{d}}\right]\right)}. \nonumber
\label{eq:GmunotZero}
\end{eqnarray}
At mid-rapidity and vanishing chemical potential,
\begin{eqnarray}
\frac{1}{2 \pi}\frac{d^2 N}{dy m_T dm_T}&& = \frac{\pm g_i V T {m_T}_i}{8 \pi^3}\times \nonumber \\ &&\frac{\varepsilon_{c,d,r}\left(\frac{-{m_T}_i}{T}\right) \; W_0\left[B\left(1-\frac{(-{m_T}_i)}{r T}\right)^{\frac{1}{d}}\right]}{(1-c)\left[1\pm\varepsilon_{c,d,r}\left(\frac{-{m_T}_i}{T}\right)\right] \left(r T+{m_T}_i\right) \left(1+W_0\left[B\left(1-\frac{(-{m_T}_i)}{r T}\right)^{\frac{1}{d}}\right]\right)}.
\end{eqnarray}
Now, we can derive the transverse momentum distribution
\begin{eqnarray}
\left.\frac{1}{2 \pi}\frac{d^2N}{p_T dy dp_T}\right|_{y=0}&&= \frac{\pm g_i V T {m_T}_i}{8 \pi^3} \times \nonumber \\ &&\frac{\varepsilon_{c,d,r}\left(\frac{-{m_T}_i}{T}\right) \; W_0\left[B\left(1-\frac{(-{m_T}_i)}{r T}\right)^{\frac{1}{d}}\right]}{(1-c)\left[1\pm\varepsilon_{c,d,r}\left(\frac{-{m_T}_i}{T}\right)\right] \left(r T+{m_T}_i\right) \left(1+W_0\left[B\left(1-\frac{(-{m_T}_i)}{r T}\right)^{\frac{1}{d}}\right]\right)}. \hspace*{8mm}
 \label{transmomentum}
\end{eqnarray}
At mid-rapidity, i.e. $y=0$ and non-vanishing chemical potential, we get
\begin{eqnarray}
\left.\frac{1}{2 \pi}\frac{d^2N}{p_T dy dp_T}\right|_{y=0}&&= \frac{\pm g_i V T {m_T}_i}{8 \pi^3} \times \nonumber \\ &&\frac{\varepsilon_{c,d,r}\left(\frac{\mu-{m_T}_i}{T}\right) \; W_0\left[B\left(1-\frac{(\mu-{m_T}_i)}{r T}\right)^{\frac{1}{d}}\right]}{(1-c)\left[1\pm\varepsilon_{c,d,r}\left(\frac{\mu-{m_T}_i}{T}\right)\right] \left(r T-\mu+{m_T}_i\right) \left(1+W_0\left[B\left(1-\frac{(\mu-{m_T}_i)}{r T}\right)^{\frac{1}{d}}\right]\right)}. \hspace*{8mm}
 \label{transmomentum1}
\end{eqnarray}

In the following section, we introduce results on the transverse momentum distributions of the strange hadrons \Kslxi measured in different high-energy experiments, at various energies \cite{Adam:2019koz,Khachatryan:2011tm,Khachatryan:2016yru} within Boltzmann and generic (non)extensive statistics.

\section{Results and Discussion}
\label{sec:res}

\subsection{Statistical fits to $p_{\mathrm{T}}$ spectra}

\begin{figure}[h]
\begin{center}
\includegraphics[scale=0.4]{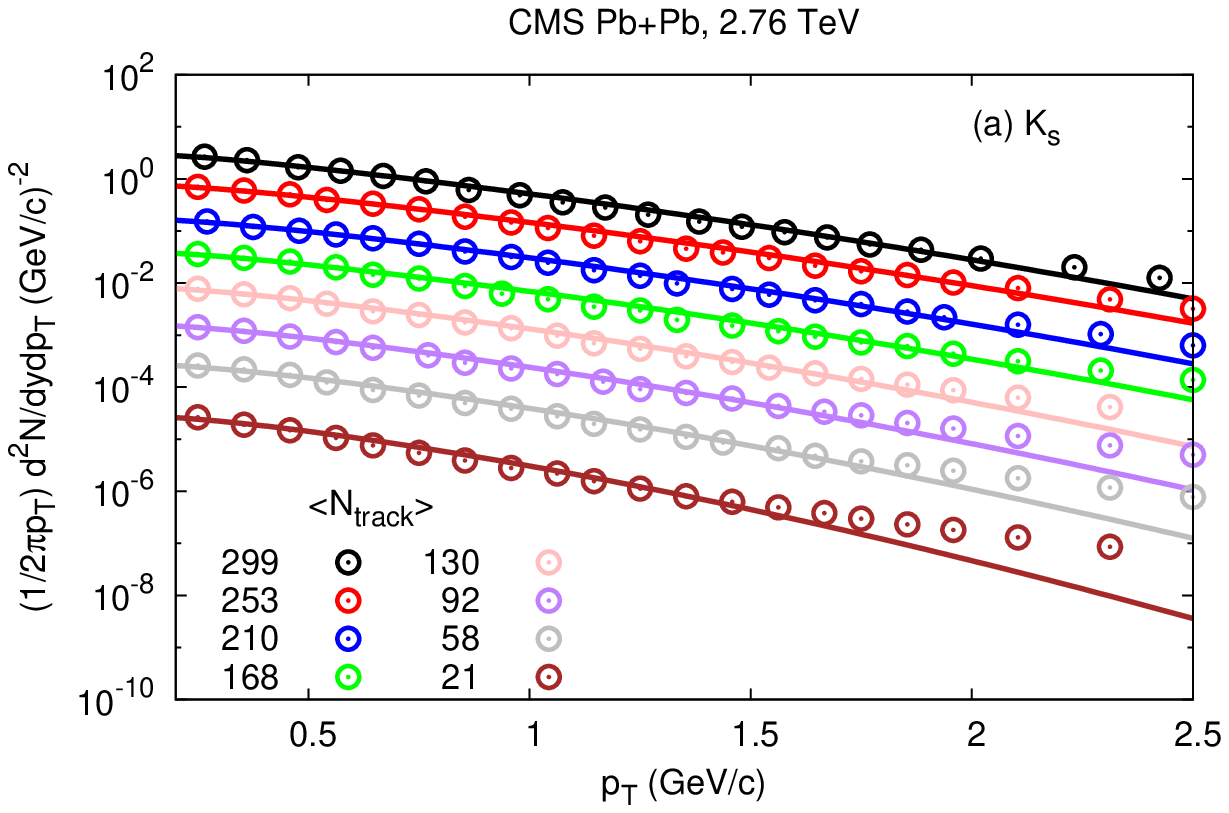}
\includegraphics[scale=0.4]{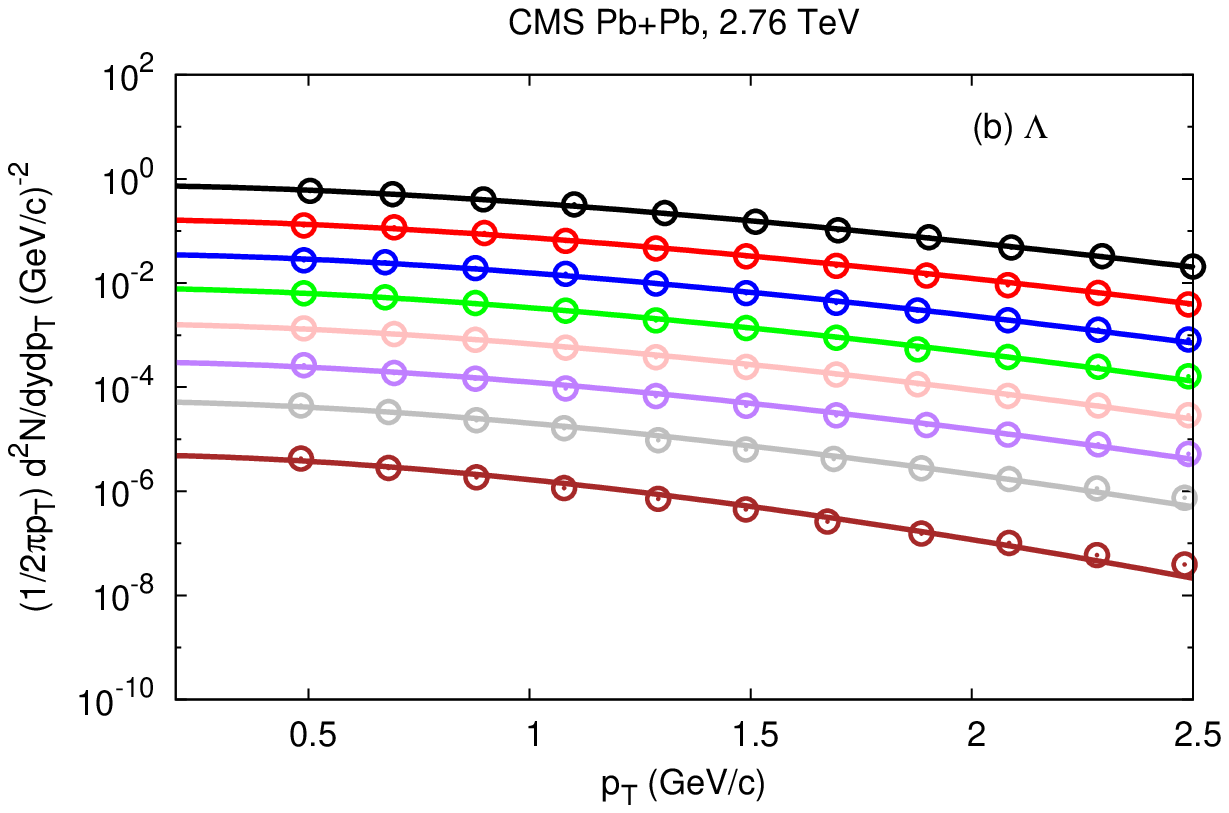}
\includegraphics[scale=0.4]{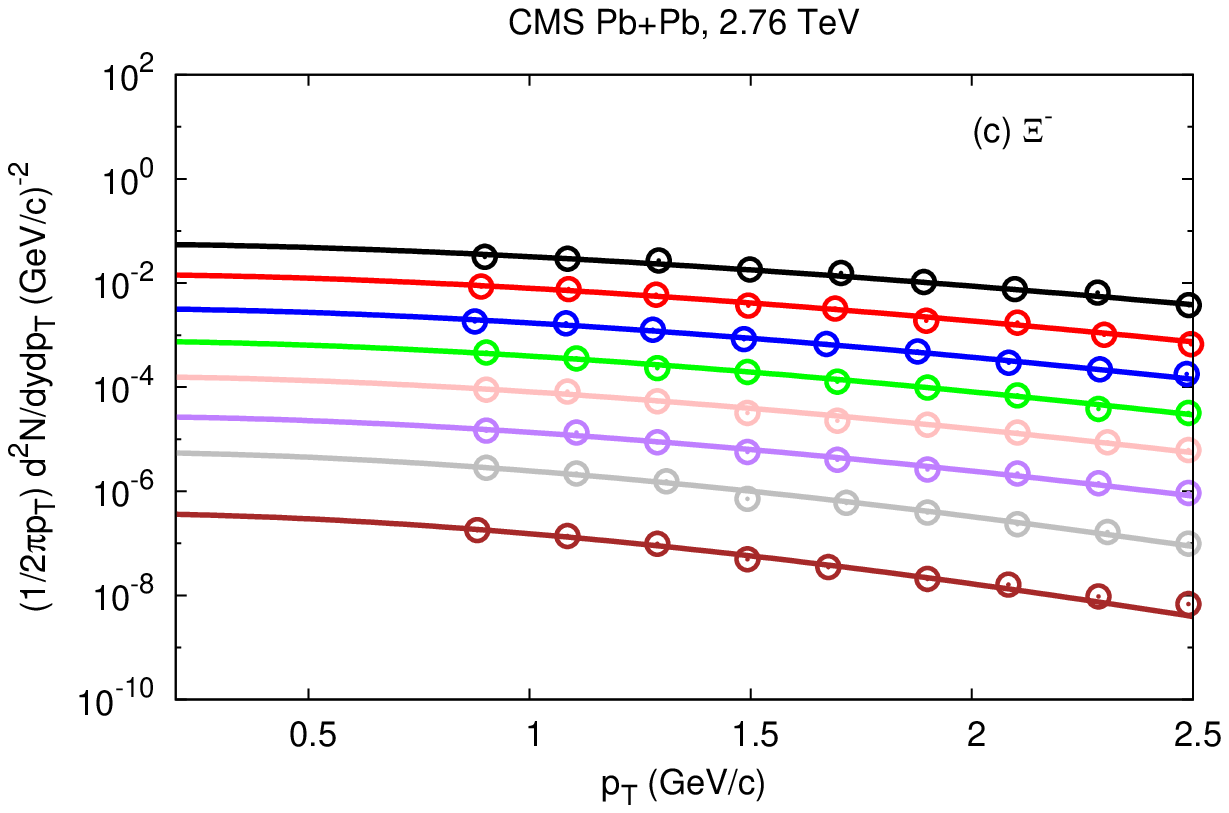}
\caption{(Color online) In different multiplicity intervals, the transverse momentum distributions of the strange hadrons \Kslxi produced in $\textsf{Pb+Pb}$ collisions, at $\sqrt{s_{\mathrm{NN}}}=2.76~$TeV (symbols) \cite{Khachatryan:2016yru} are fitted to generic (non)extensive statistical approaches (solid curves), Eq.~\ref{transmomentum}. \label{fit:2.76} }
\end{center}
\end{figure}

\begin{figure}[h]
\begin{center}
\includegraphics[scale=0.4]{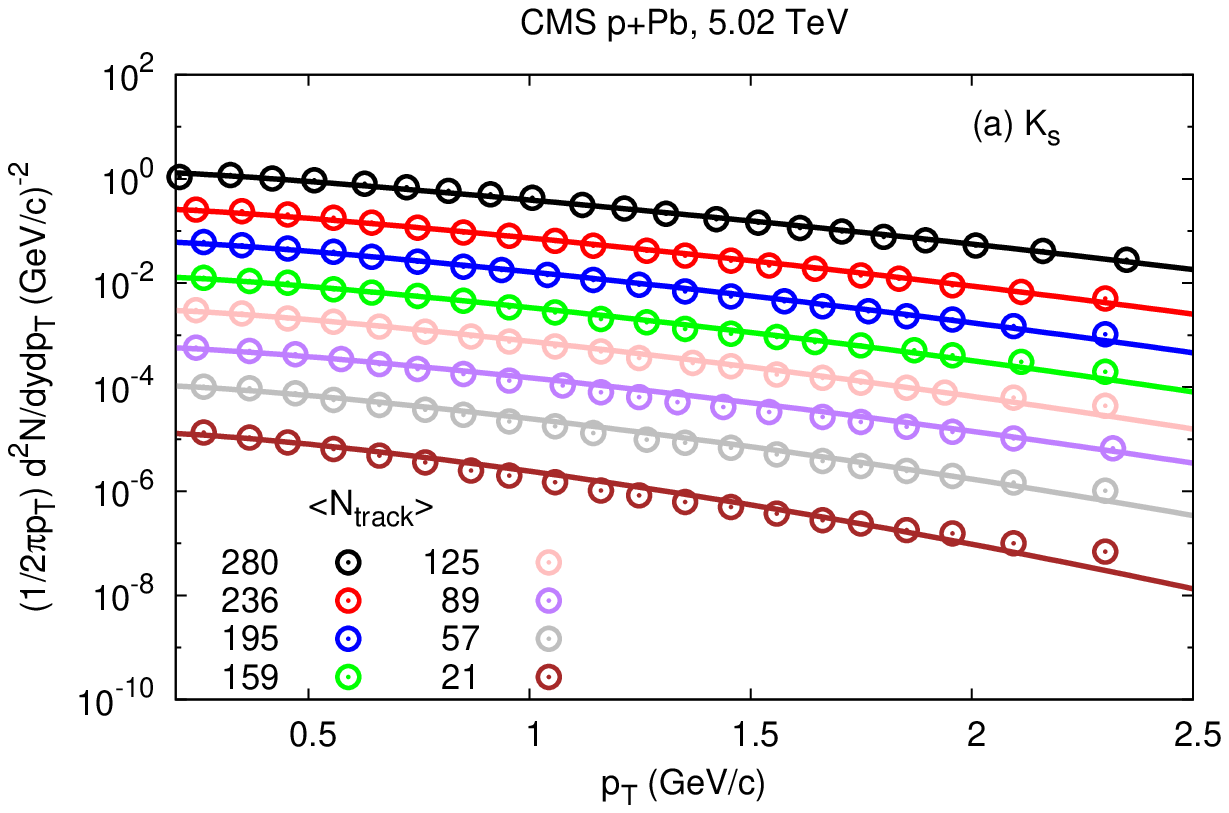}
\includegraphics[scale=0.4]{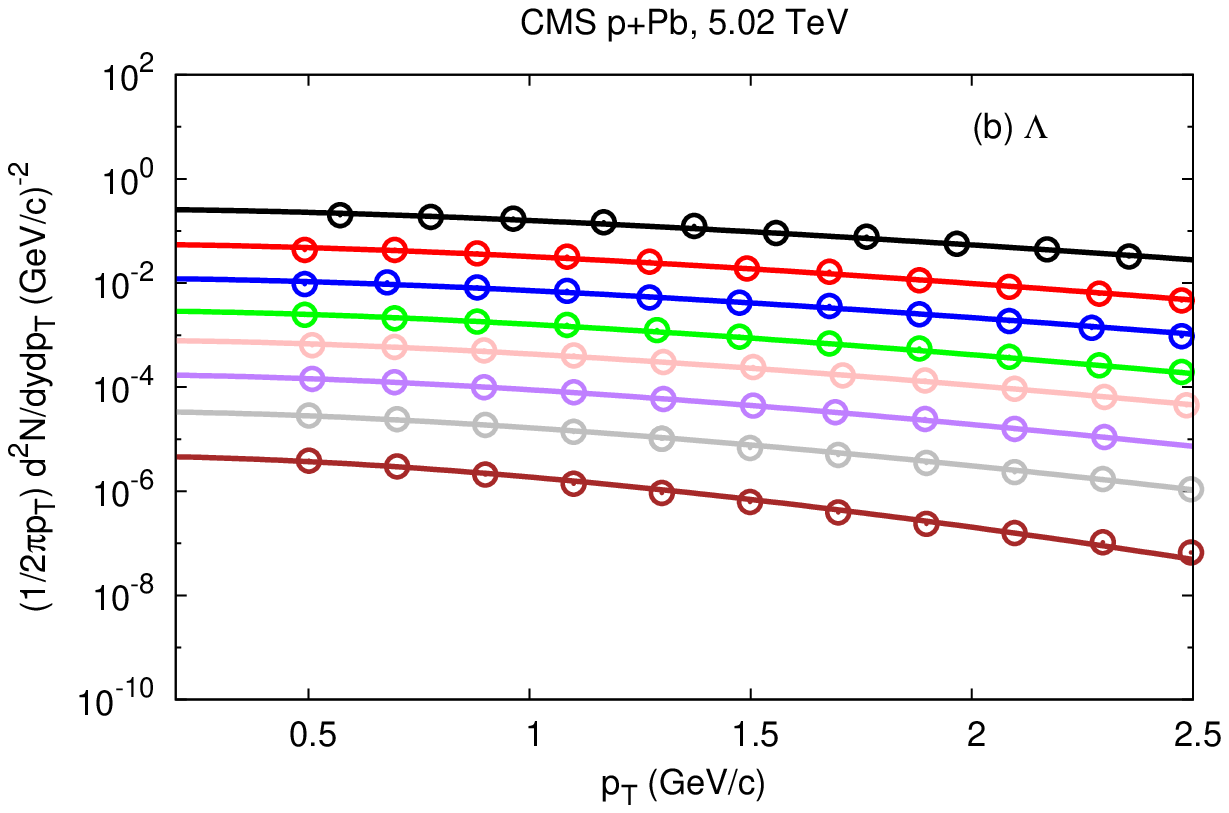}
\includegraphics[scale=0.4]{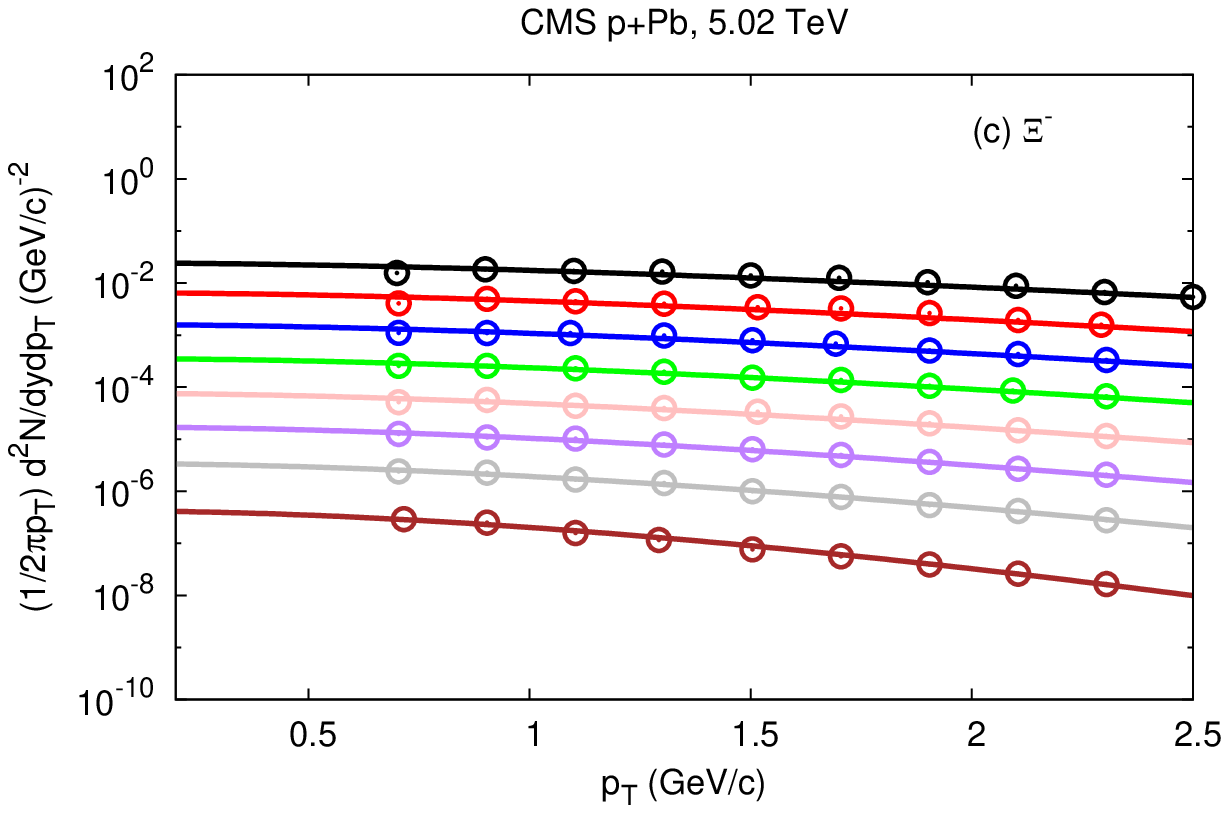}
\caption{Similar to Fig. \ref{fit:2.76}, but in $\textsf{p+Pb}$ collisions, at $\sqrt{s_{\mathrm{NN}}}=5.02~$TeV.}
\label{fit:5.02}
\end{center}
\end{figure}

\begin{figure}[t!]
\begin{center}
\includegraphics[scale=0.4]{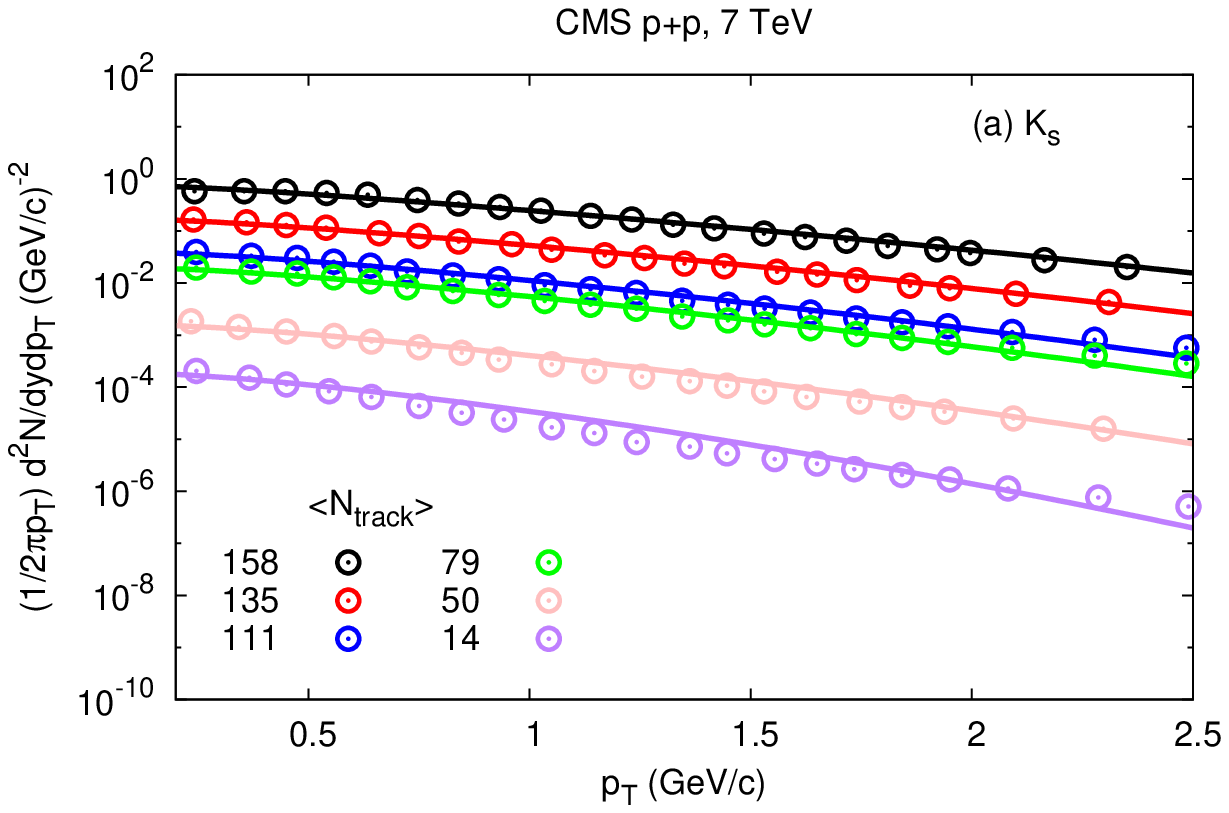}
\includegraphics[scale=0.4]{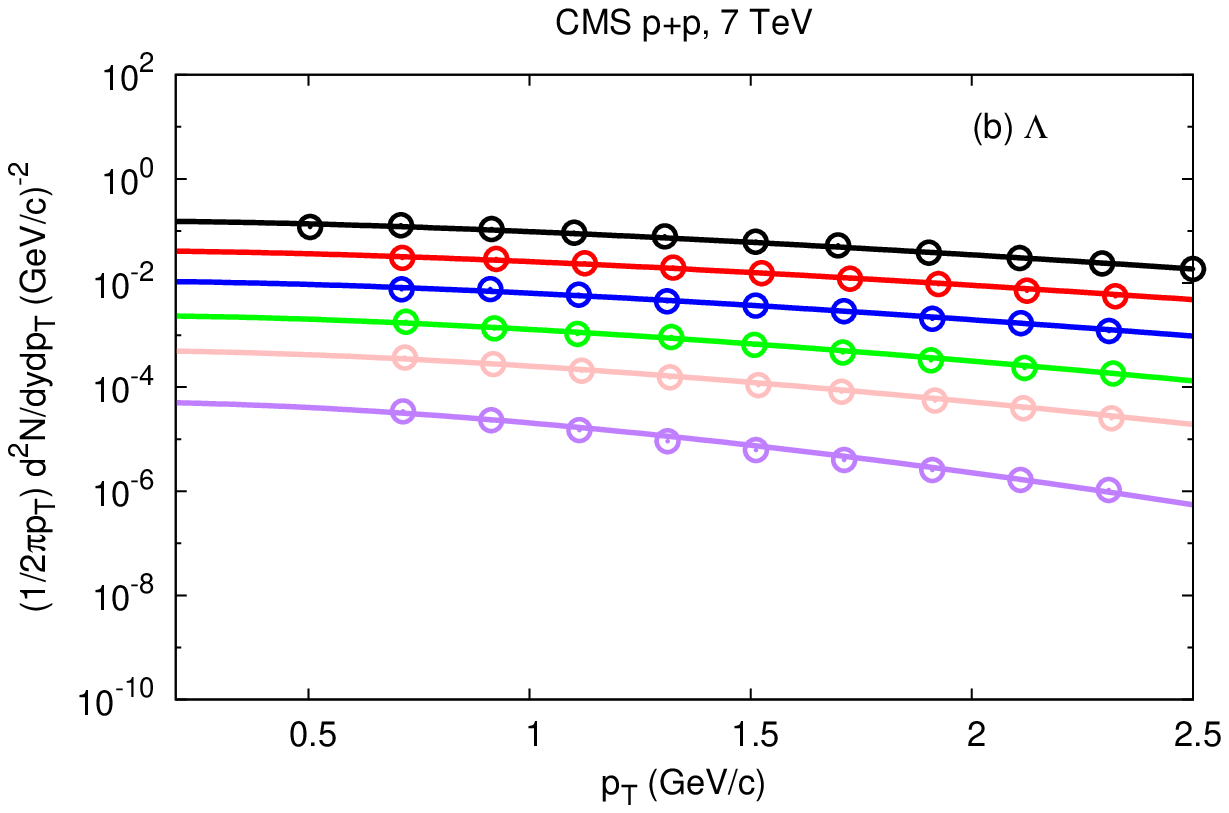}
\includegraphics[scale=0.4]{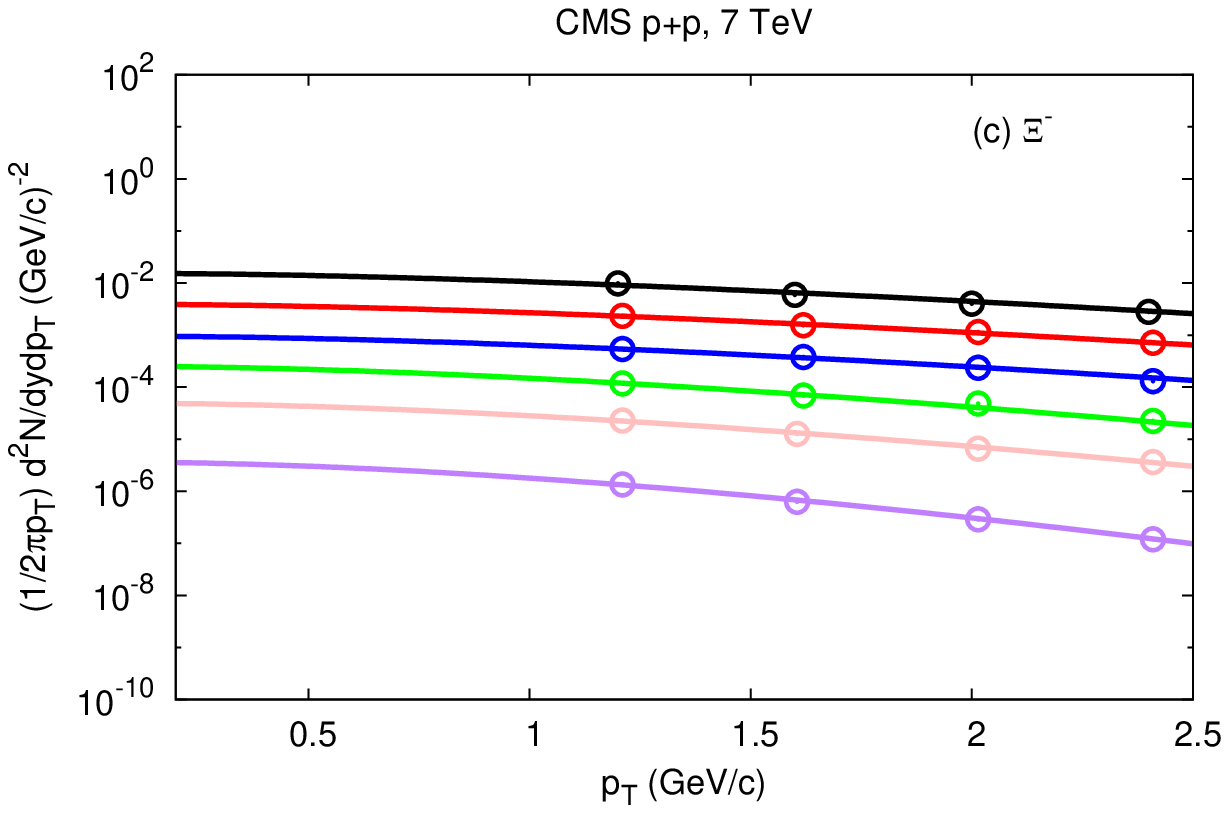}
\caption{The same as in Fig. \ref{fit:2.76}, but in $\textsf{p+p}$ collisions, at $\sqrt{s_{\mathrm{NN}}}=7~$TeV. \label{fit:7}}
\end{center}
\end{figure}

The transverse momentum spectra of the strange hadrons \Kslxi produced in $\textsf{Pb+Pb}$ collisions, at $\sqrt{s_{\mathrm{NN}}}=2.76~$TeV, $\textsf{p+Pb}$ collisions, at $\sqrt{s_{\mathrm{NN}}}= 5.02~$TeV, and $\textsf{p+p}$ collisions, at $\sqrt{s_{\mathrm{NN}}}=0.2$, $0.9$, $7~$TeV \cite{Khachatryan:2016yru} in different multiplicity intervals are fitted using generic (non)extensive statistics and shown in Figs. \ref{fit:2.76}, \ref{fit:5.02}, and \ref{fit:7}, respectively. The goodness of the statistical fits are listed in Tab. \ref{Tab1}.

Figure~\ref{fit:2.76} presents the transverse momentum $p_{\mathrm{T}}$ spectra of (a) $K_s^0$, (b) $\Lambda$, and (c) $\Xi^-$, produced in $\textsf{Pb+Pb}$ collision, at $\sqrt{s_{\mathrm{NN}}}= 2.76~$TeV. Symbols refer to the experimental data of the \textit{CMS} experiment \cite{Khachatryan:2016yru}, which are divided into multiplicity intervals $N_{\mathrm{trk}}^{\mathrm{offline}}$ in the mid-rapidity range $\vert y \vert<1.0$. The corresponding averaged multiplicity are $<N_{\mathrm{track}}>=21$, $58$, $92$, $130$, $168$, $210$, $253$, and $299$ \cite{Chatrchyan:2013nka}. Our calculations based on generic (non)extensive statistics are represented by solid curves. Here, we are interested on the smallest $p_{\mathrm{T}}$ region. It is found that for all multiplicity intervals our calculations for $\Lambda$ and $\Xi^-$ hadrons agree well with the experimental data of $\textsf{Pb+Pb}$ collisions, at $\sqrt{s_{\mathrm{NN}}}=2.76~$TeV. For $K_s^0$, the agreement becomes worse, specially with decreasing multiplicity, i.e. moving towards peripherality.

Figure~\ref{fit:5.02} depicts the transverse momentum distribution $p_{\mathrm{T}}$ of (a) $K_s^0$, (b) $\Lambda$, and (c) $\Xi^-$, produced in $\textsf{p+Pb}$ collisions, at $\sqrt{s_{\mathrm{NN}}}=5.02~$TeV in different multiplicity intervals. In the mid-rapidity range $\vert y \vert<1.0$, the experimental data (symbols) are divided into different multiplicity intervals $N_{\mathrm{trk}}^{\mathrm{offline}}=21$, $57$, $89$, $125$, $159$, $195$, $236$ and $280$ \cite{Chatrchyan:2013nka,Khachatryan:2016yru}. The solid curves are the results calculated using generic (non)extensive statistics. We notice that for all multiplicity intervals our results for all hadrons are in a good agreement with the experimental data of $\textsf{p+Pb}$ collisions, at $\sqrt{s_{\mathrm{NN}}}= 5.02~$TeV.

For different multiplicity intervals, Fig.~\ref{fit:7} presents the transverse momentum distribution $p_{\mathrm{T}}$ spectra of (a) $K_s^0$, (b) $\Lambda$, and (c) $\Xi^-$ produced in $\textsf{p+p}$ collisions, at $\sqrt{s_{\mathrm{NN}}}=7~$TeV. The \textit{CMS} results (symbols) divided into $N_{\mathrm{trk}}^{\mathrm{offline}}=14$, $50$, $79$, $111$, $135$ and $158$ \cite{Chatrchyan:2013nka,Khachatryan:2016yru} in the mid-rapidity range $\vert y \vert<1.0$ are compared with our calculations (curves). We observe that for all multiplicity intervals and for all hadrons our calculations are in good agreement with the results measured in the $\textsf{p+p}$ collisions, at $\sqrt{s_{\mathrm{NN}}}=7~$TeV.

\begin{figure}[h]
\begin{center}
\includegraphics[scale=0.4]{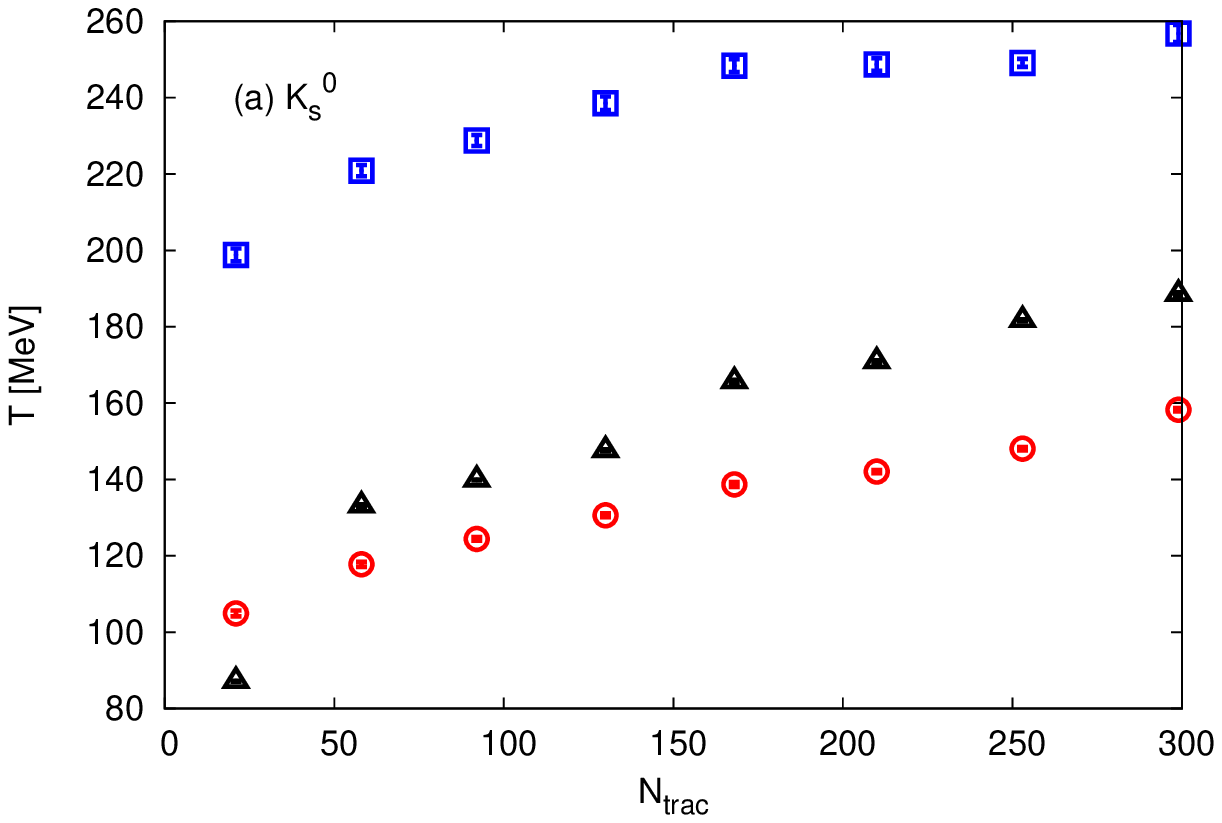}
\includegraphics[scale=0.4]{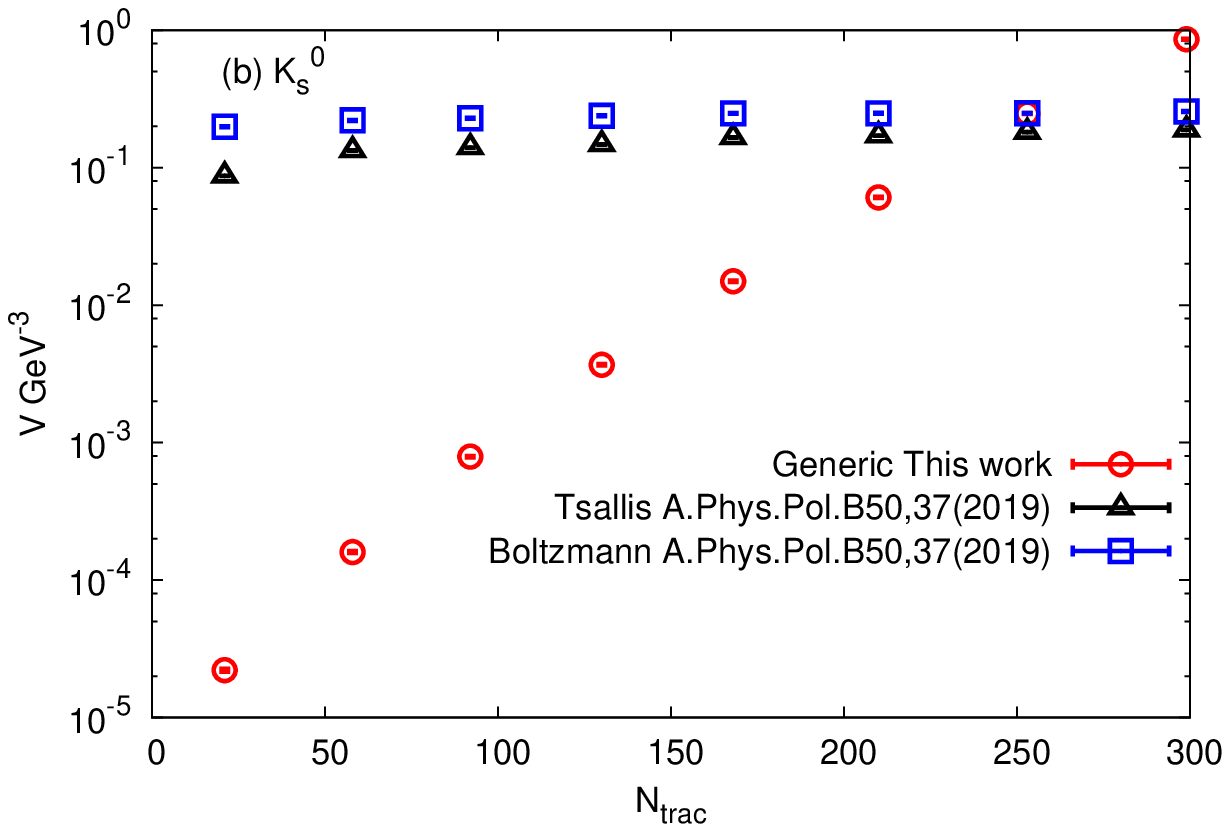}\\
\includegraphics[scale=0.4]{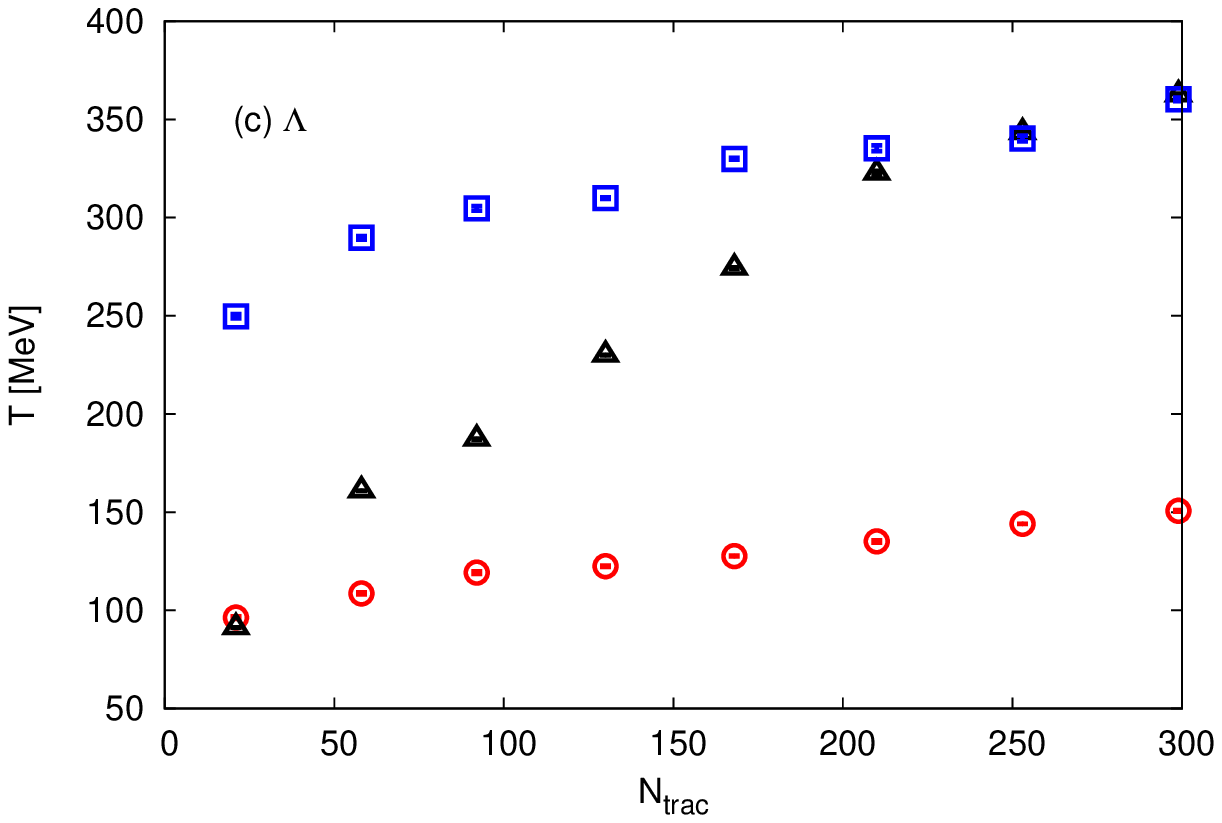}
\includegraphics[scale=0.4]{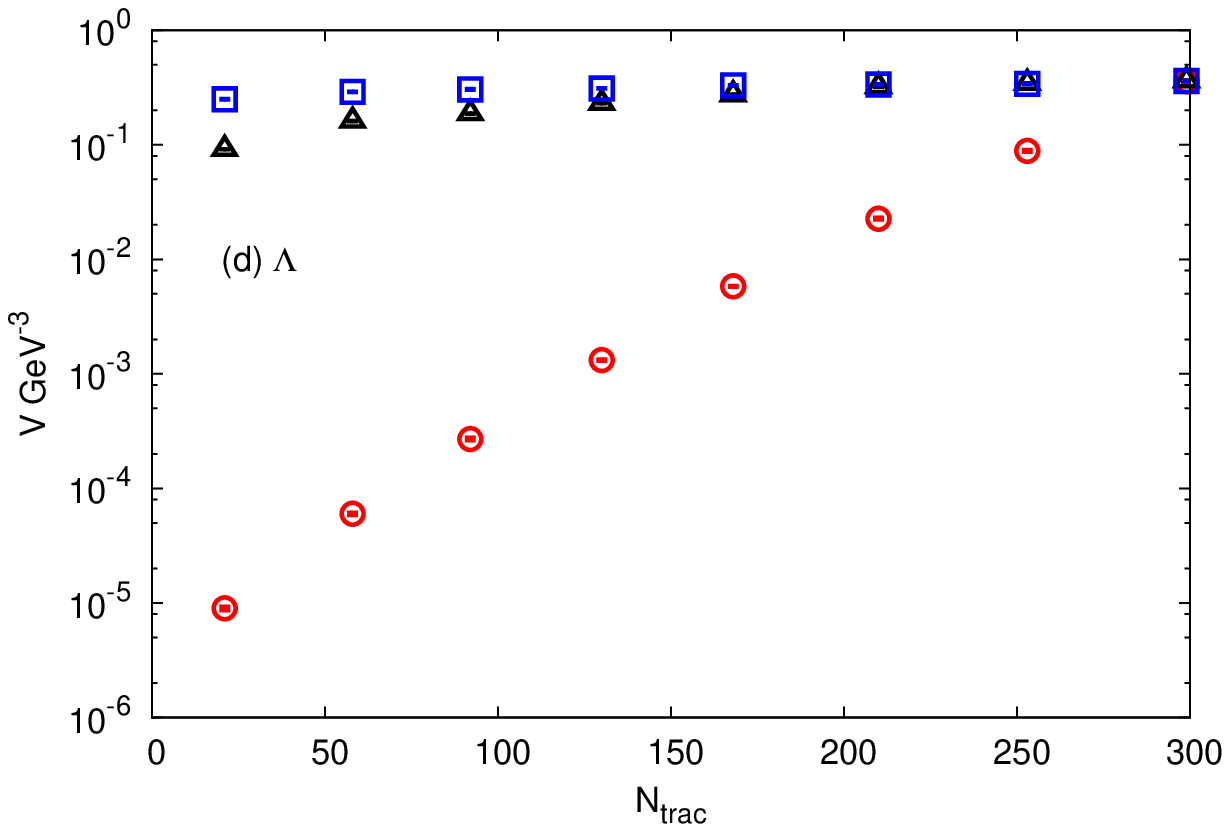}\\
\includegraphics[scale=0.4]{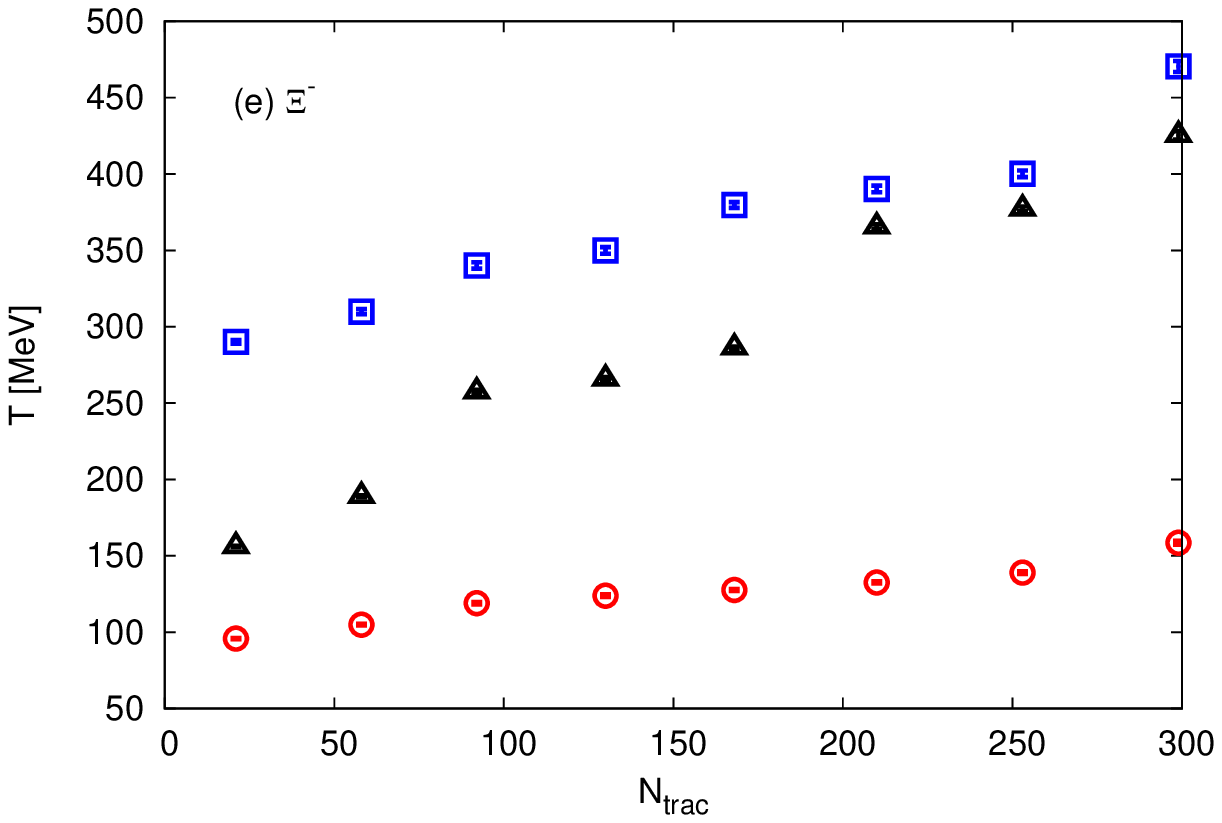}
\includegraphics[scale=0.4]{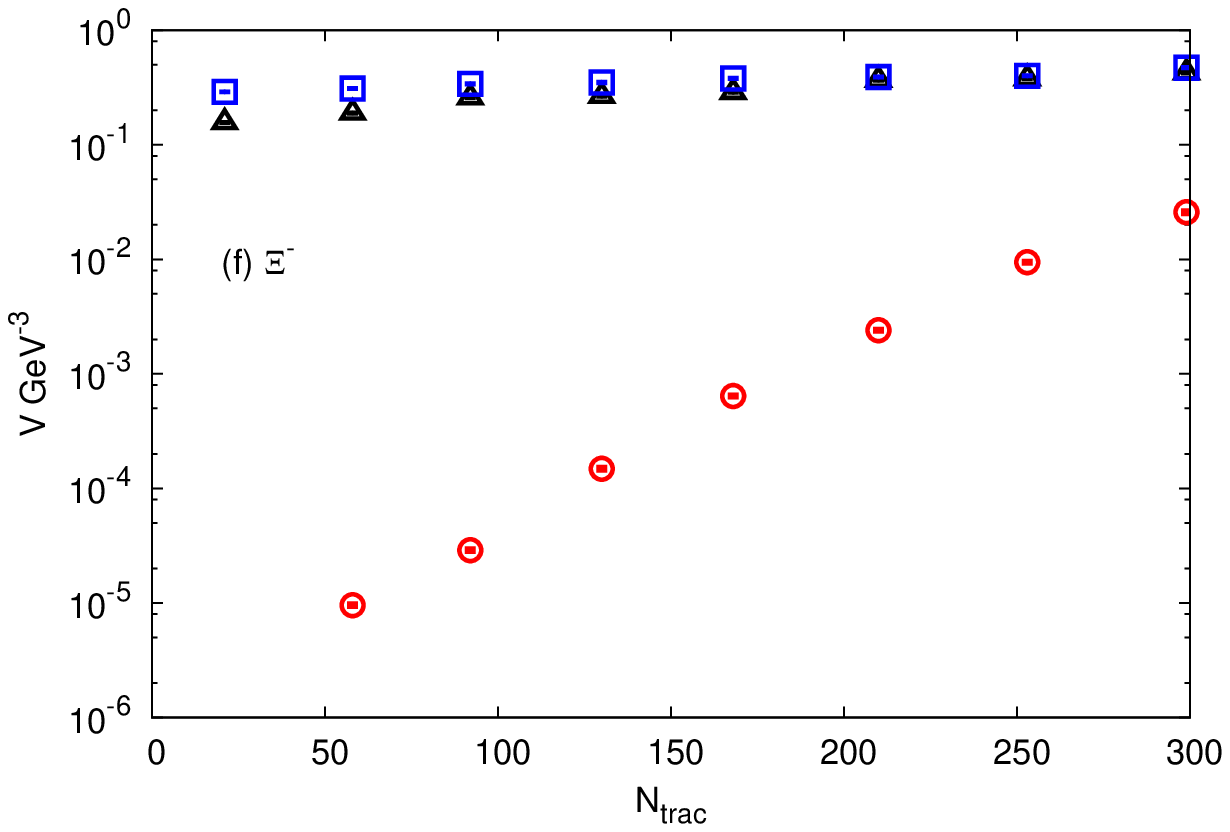}
\caption{(Color online) The resulting generic (non)extensive fit parameters $T$ and $V$ deduced from the strange hadrons \Kslxi produced in $\textsf{Pb+Pb}$ collisions as functions of multiplicity intervals, at $\sqrt{s_{\mathrm{NN}}}=2.76~$TeV, are compared with Boltzmann and Tsallis statistics \cite{Yassin:2018svv}.
\label{parameters_Ntrack_PbPb}}
\end{center}
\end{figure}

\begin{figure}[h]
\begin{center}
\includegraphics[scale=0.4]{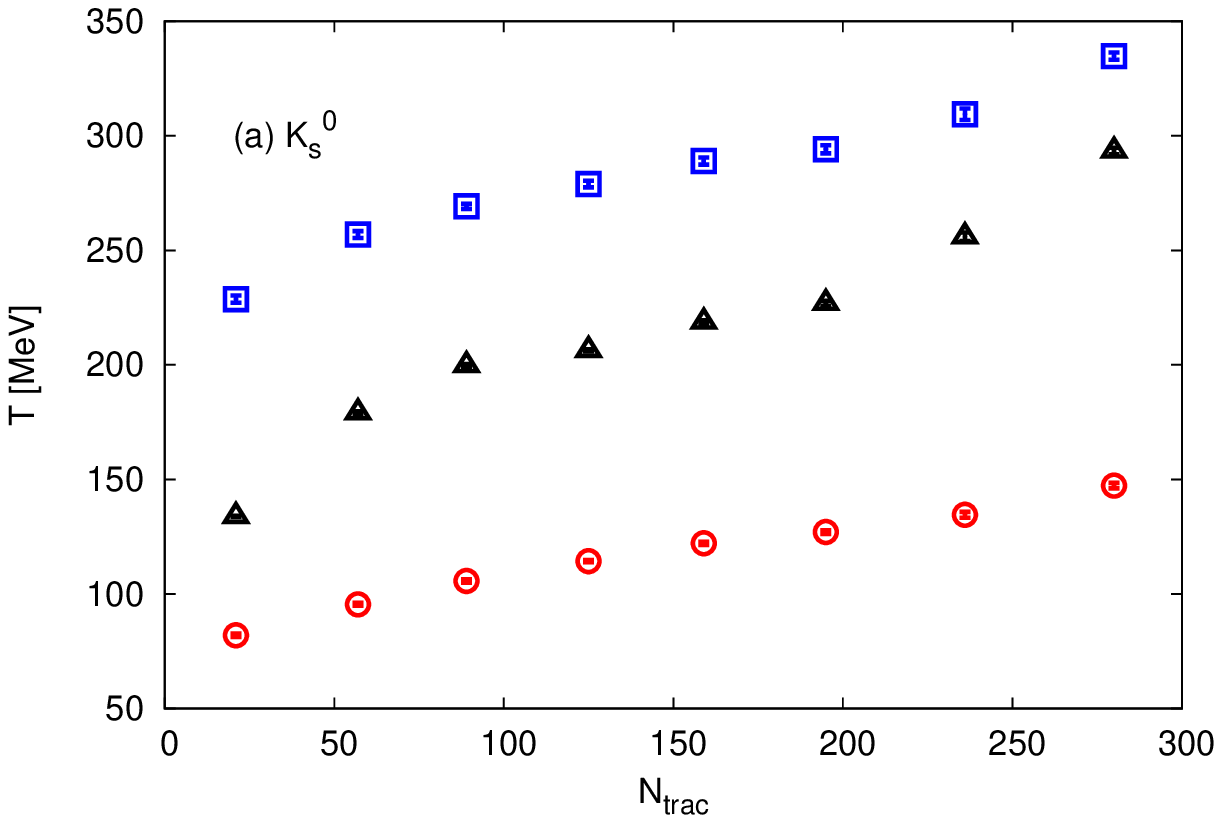}
\includegraphics[scale=0.4]{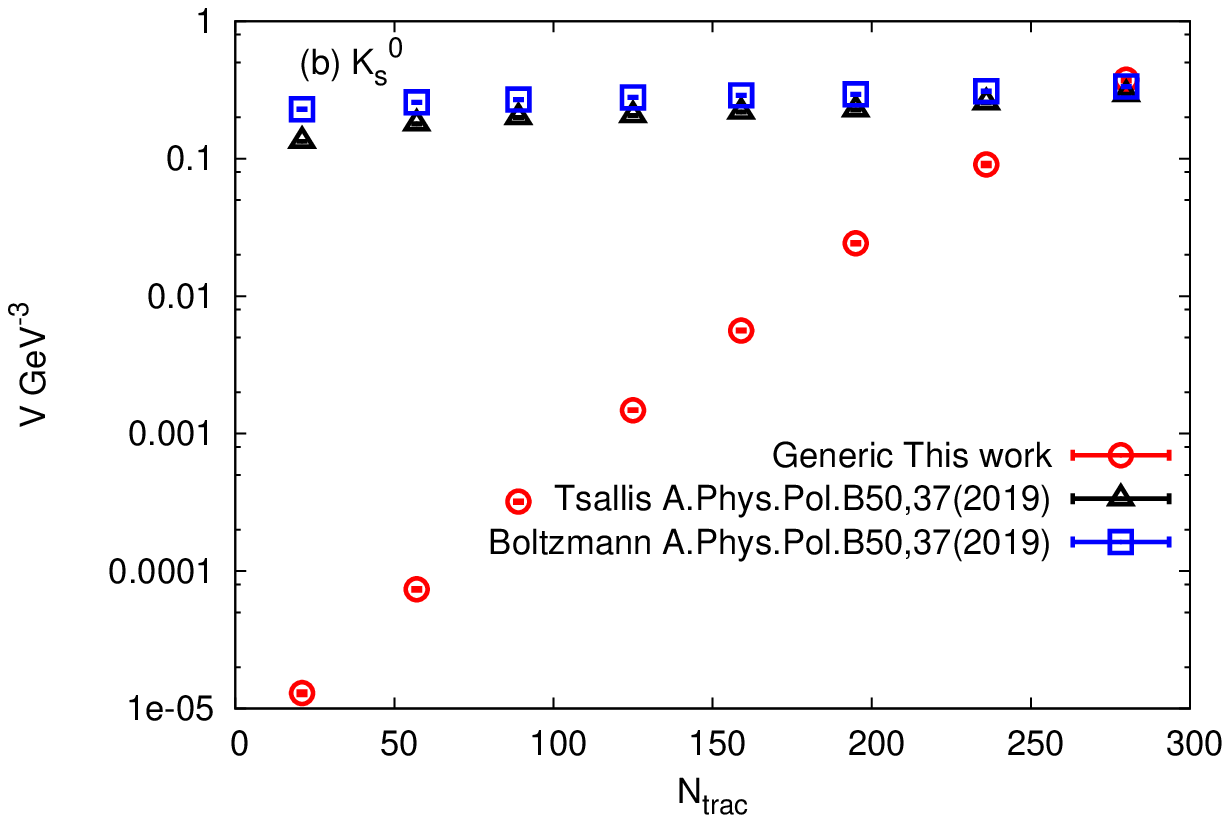}\\
\includegraphics[scale=0.4]{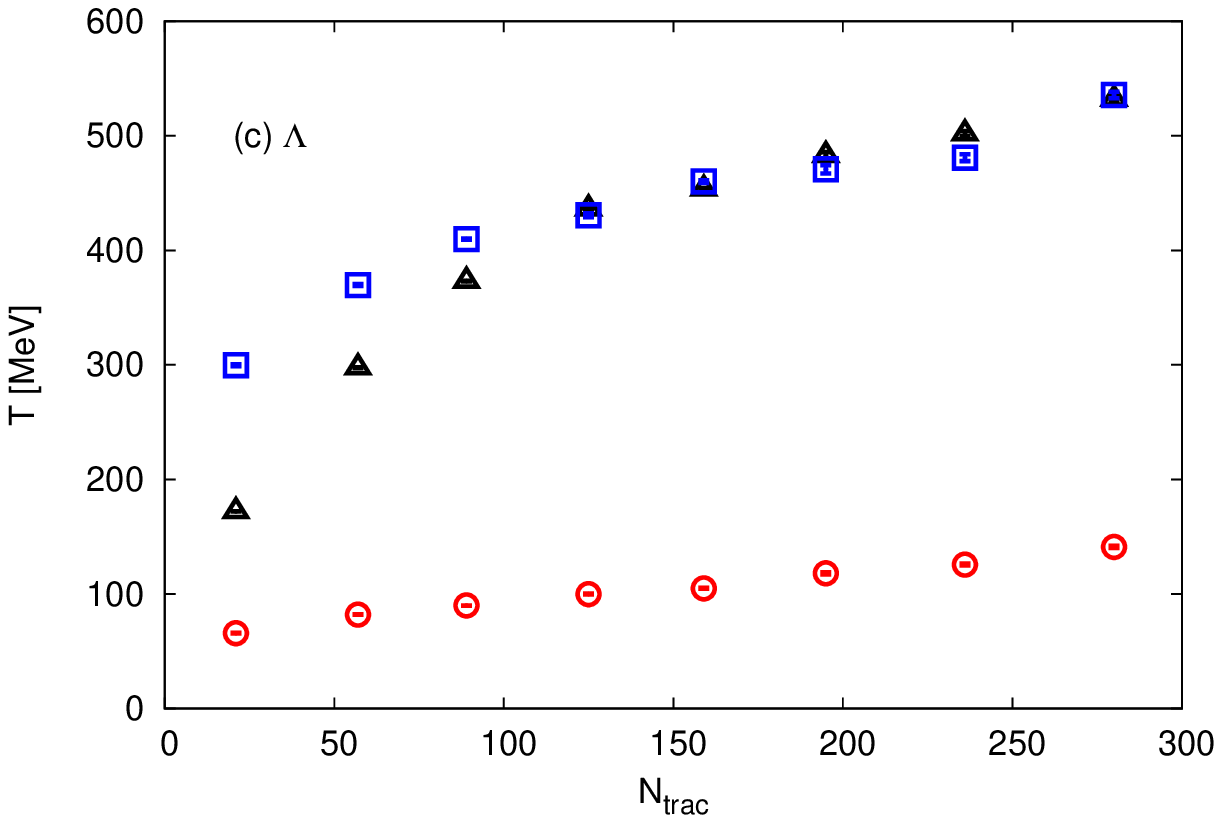}
\includegraphics[scale=0.4]{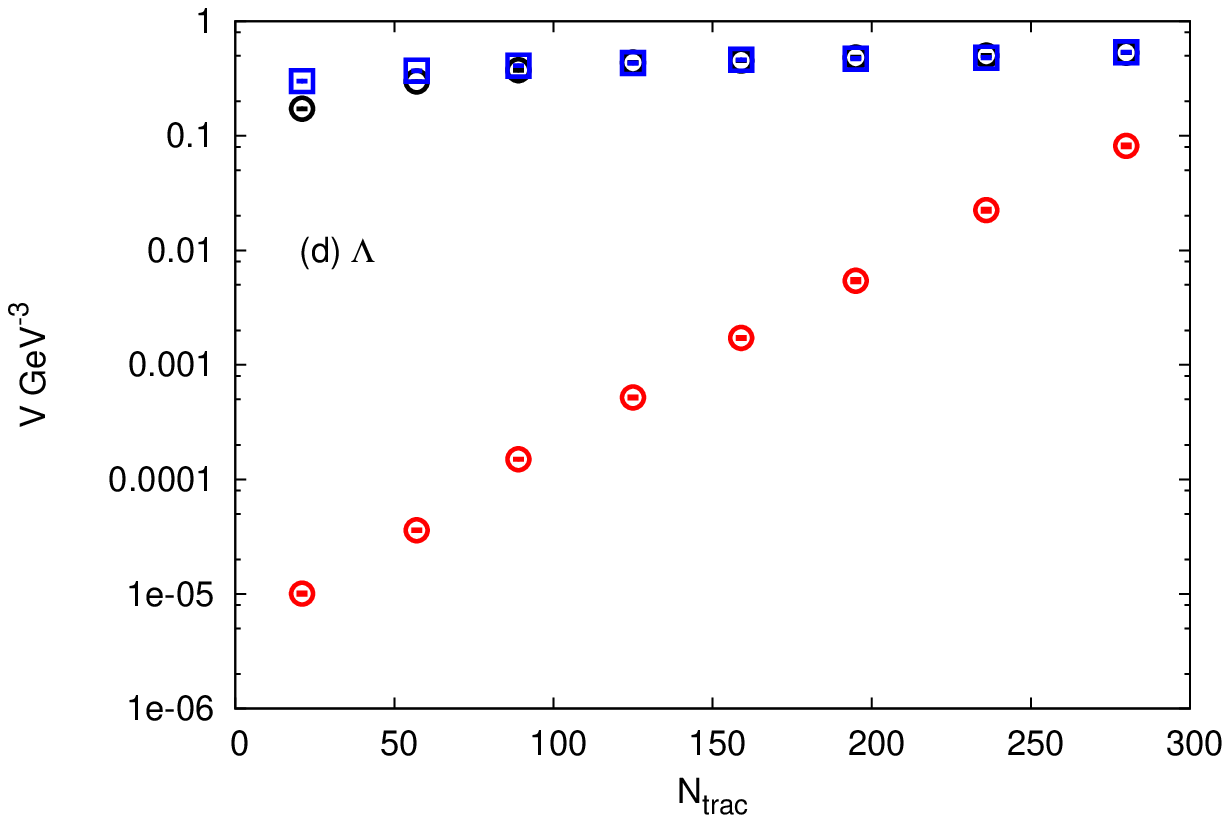}\\
\includegraphics[scale=0.4]{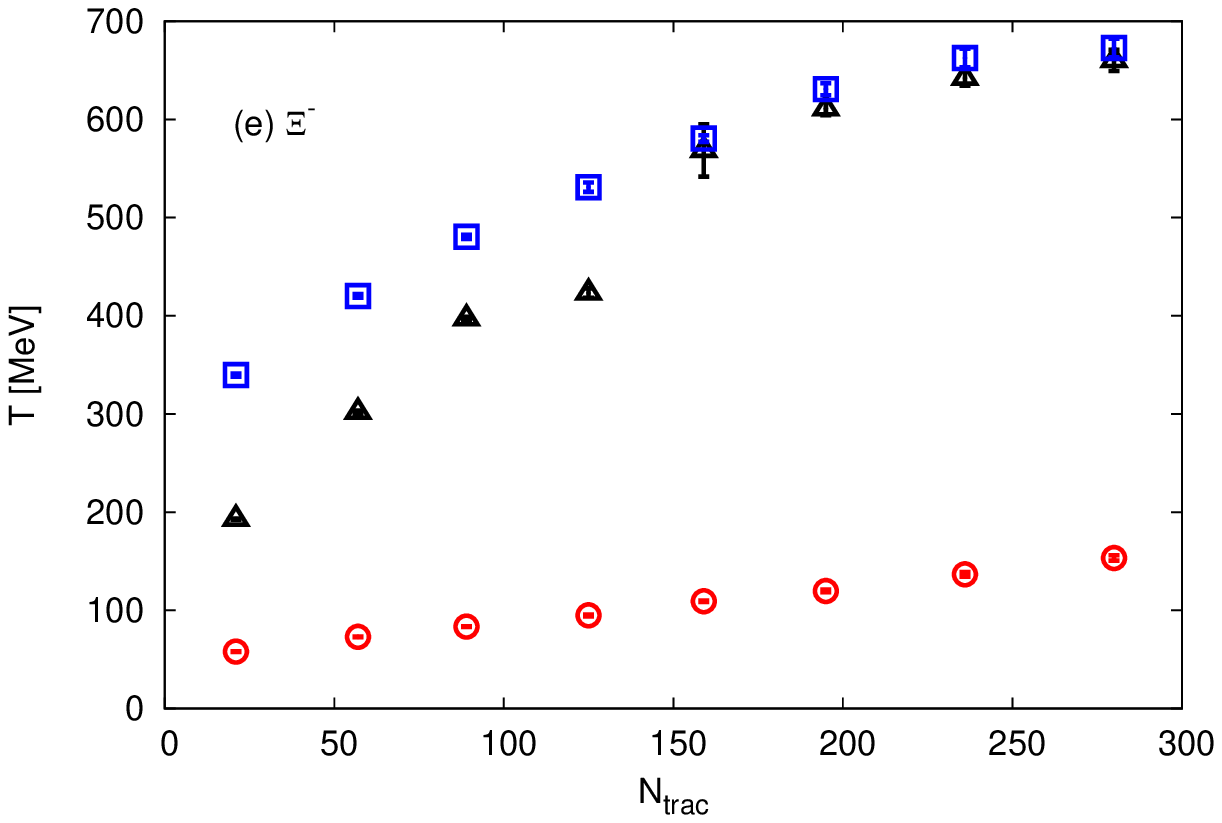}
\includegraphics[scale=0.4]{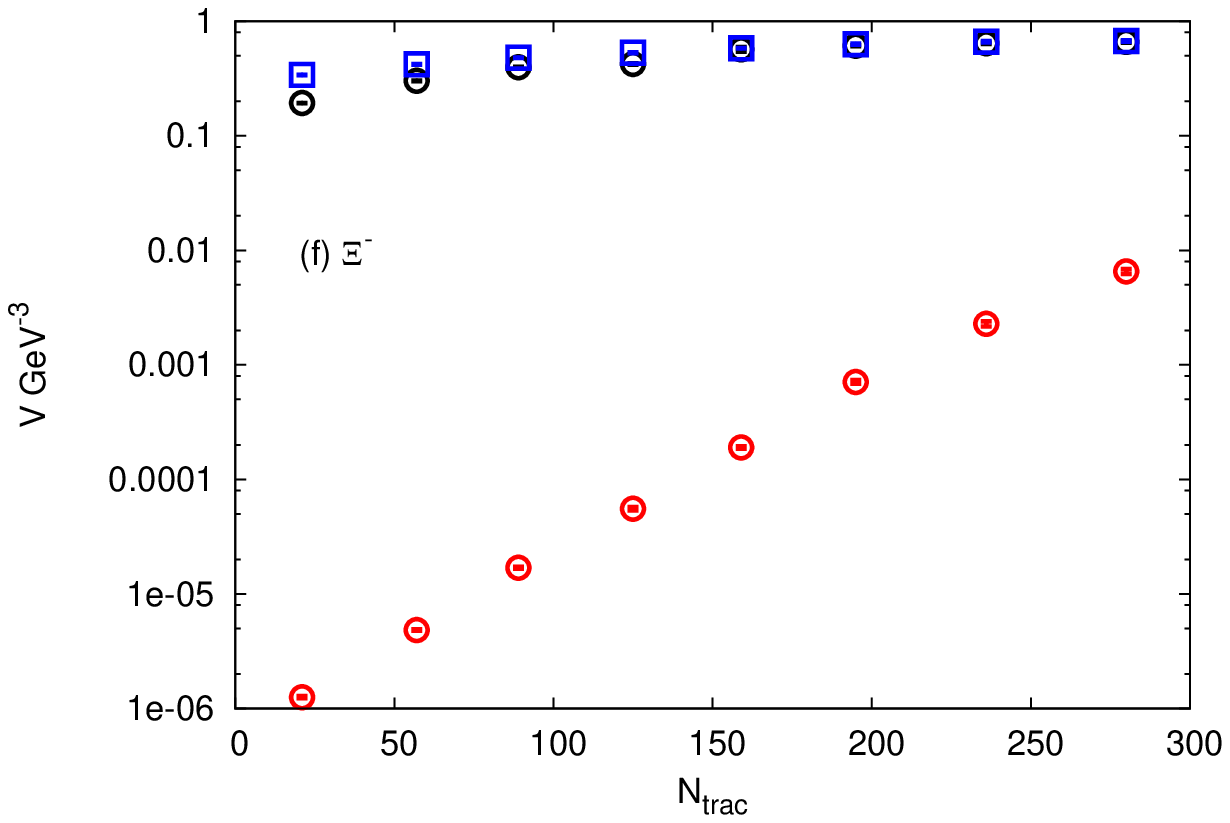}
\caption{(Color online) The same as in Fig. \ref{parameters_Ntrack_PbPb}, but here from $\textsf{p+Pb}$ collisions, at $\sqrt{s_{\mathrm{NN}}}=5.02~$TeV.
\label{parameters_Ntrack_pPb}}
\end{center}
\end{figure}

\begin{figure}[h]
\begin{center}
\includegraphics[scale=0.4]{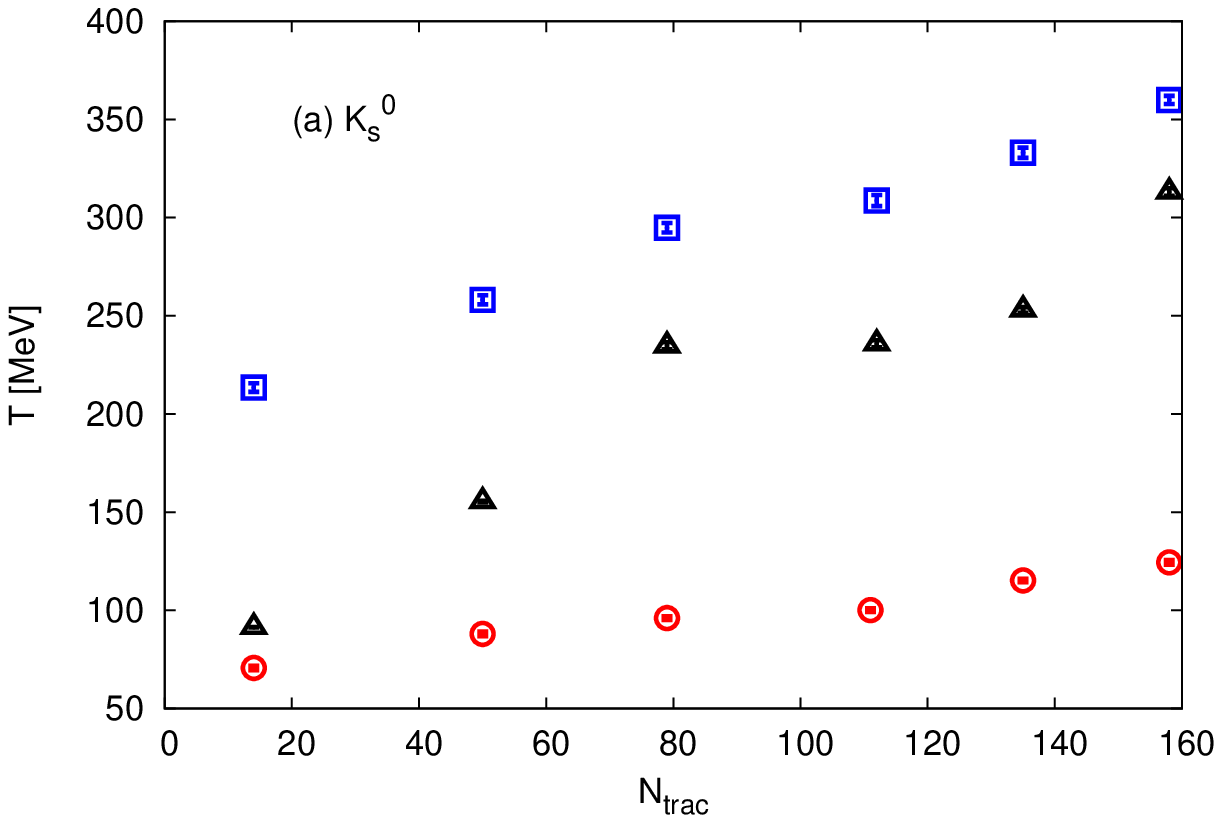}
\includegraphics[scale=0.4]{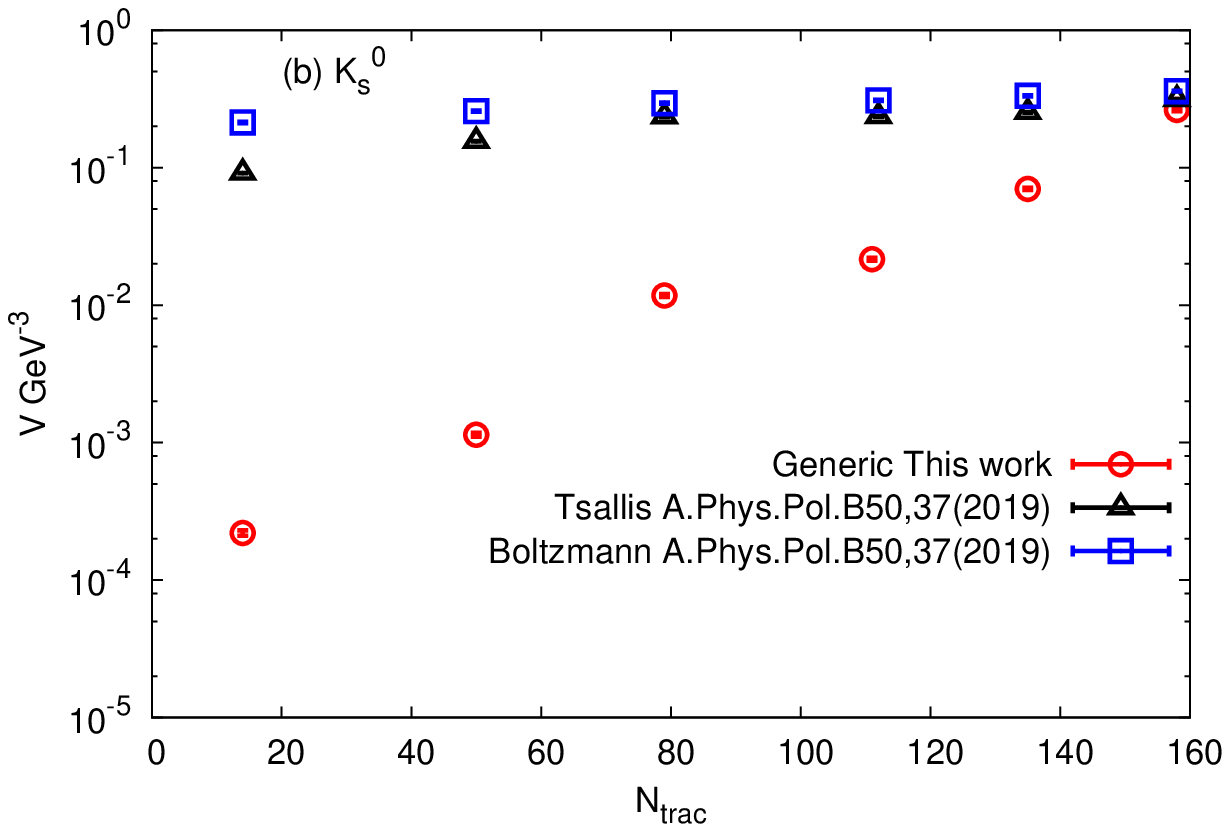}\\
\includegraphics[scale=0.4]{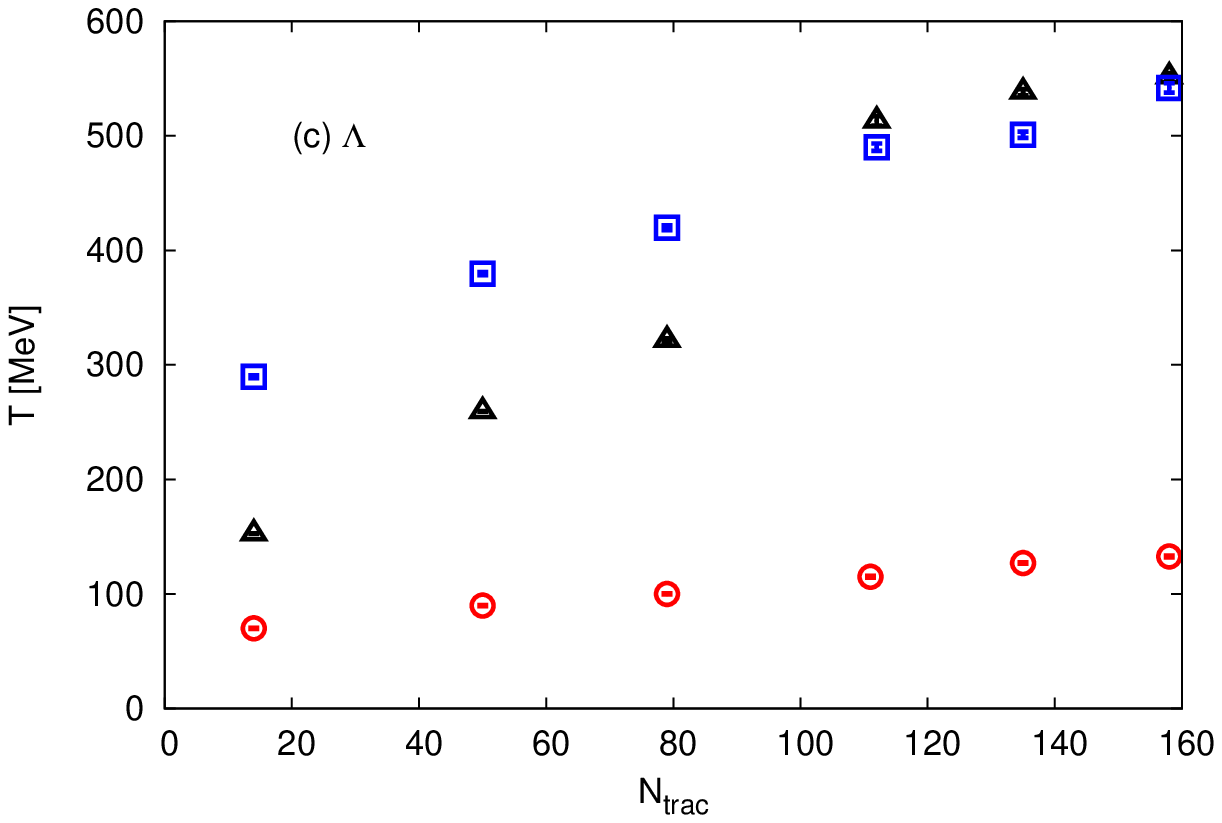}
\includegraphics[scale=0.4]{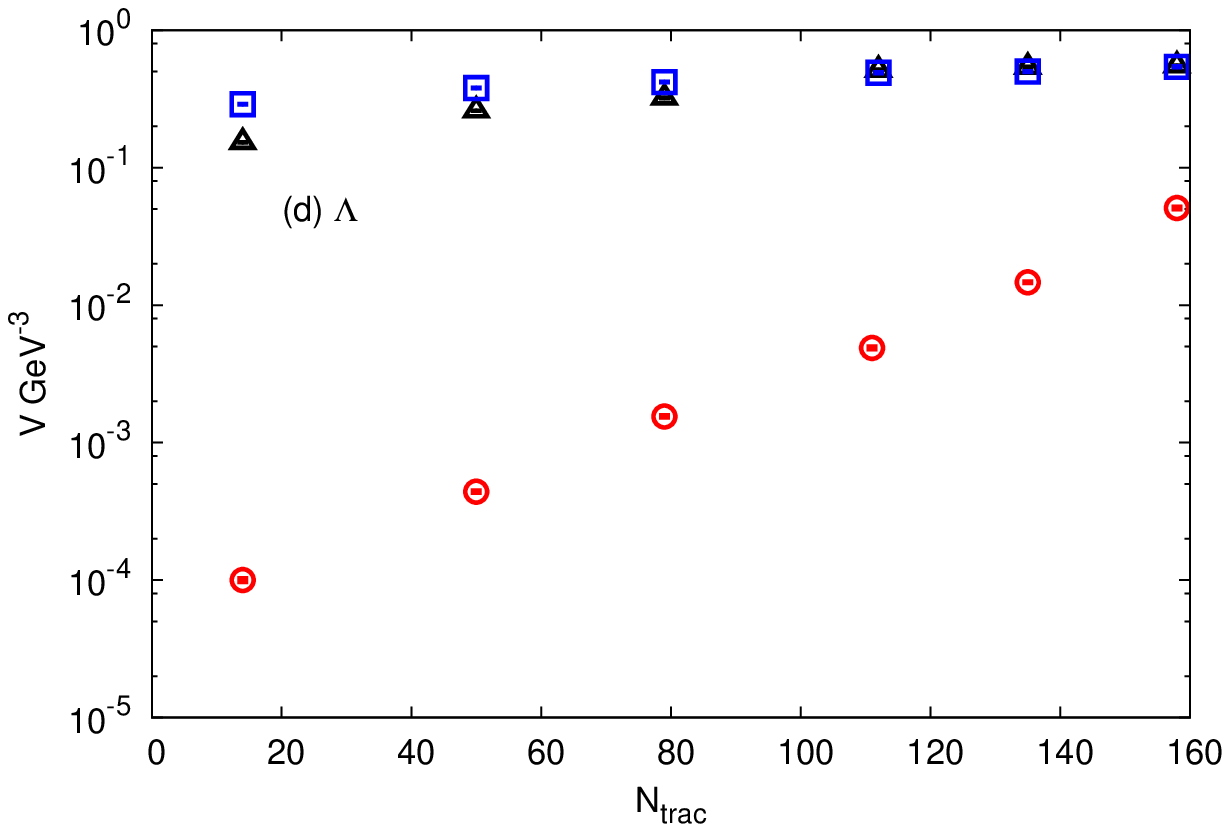}\\
\includegraphics[scale=0.4]{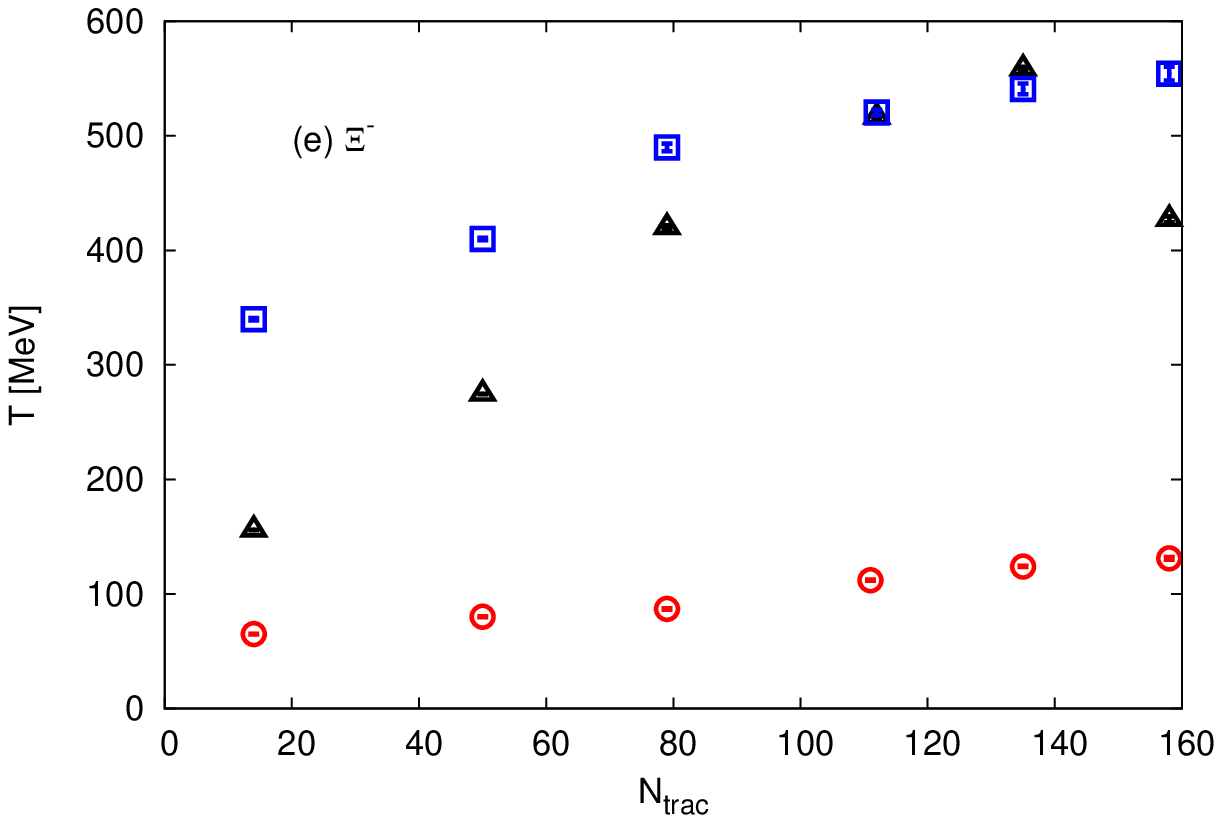}
\includegraphics[scale=0.4]{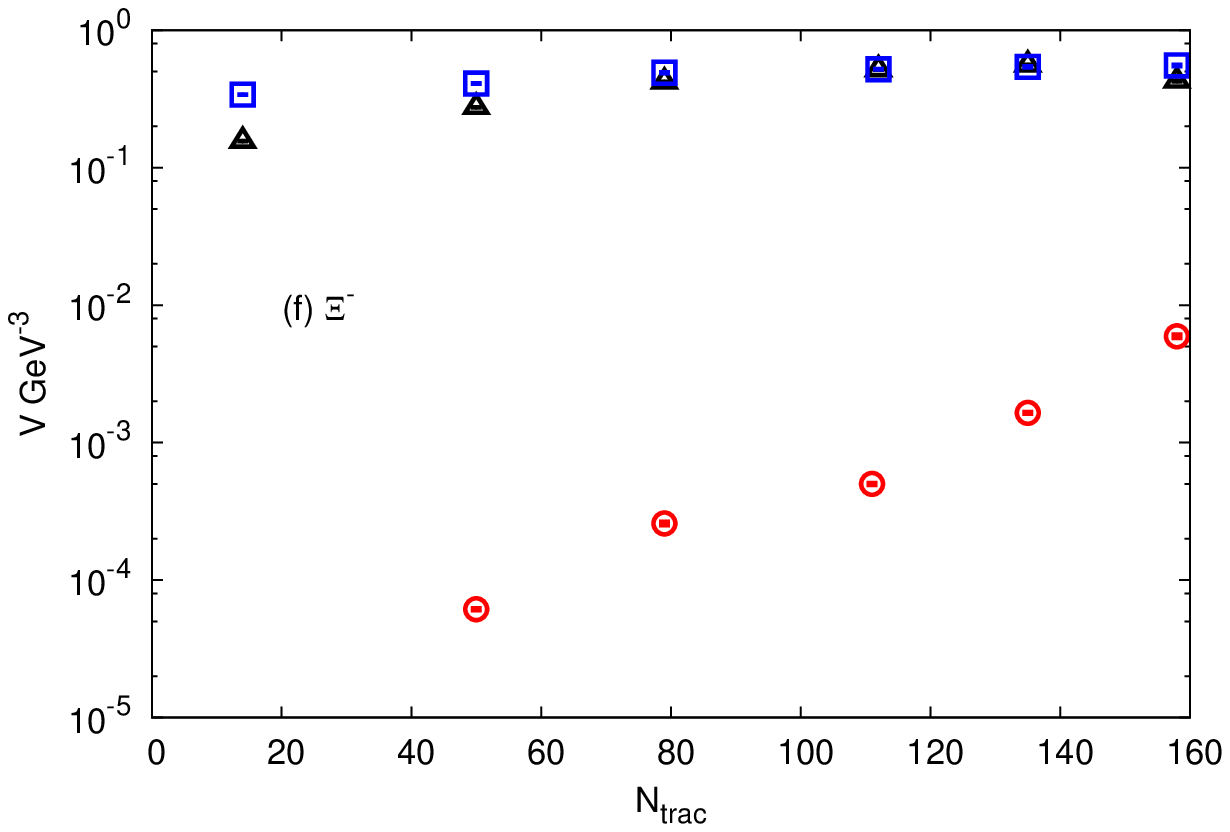}
\caption{(Color online) The same as in Fig. \ref{parameters_Ntrack_PbPb}, but here from $\textsf{p+p}$ collisions, at $\sqrt{s_{\mathrm{NN}}}=7~$TeV.}
\label{parameters_Ntrack_pp}
\end{center}
\end{figure}

\begin{figure}[h]
\begin{center}
\includegraphics[scale=0.4]{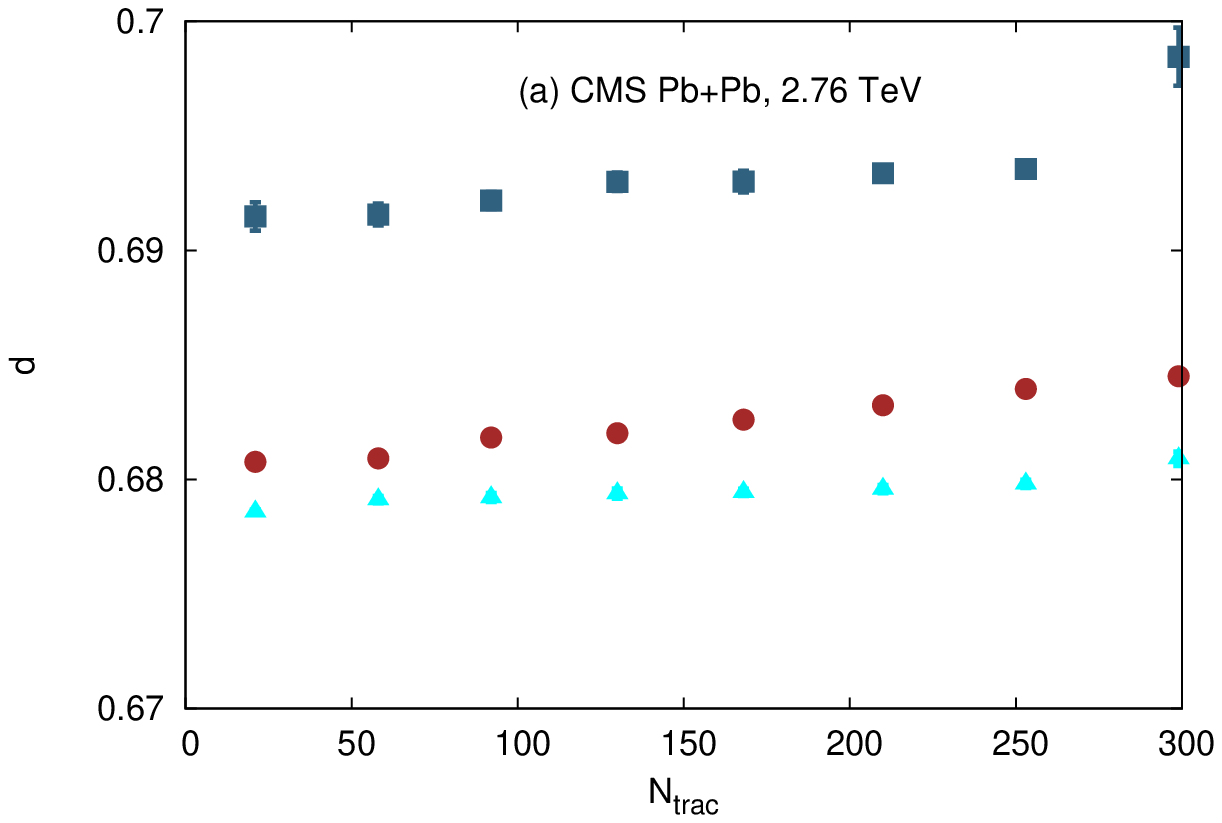}
\includegraphics[scale=0.4]{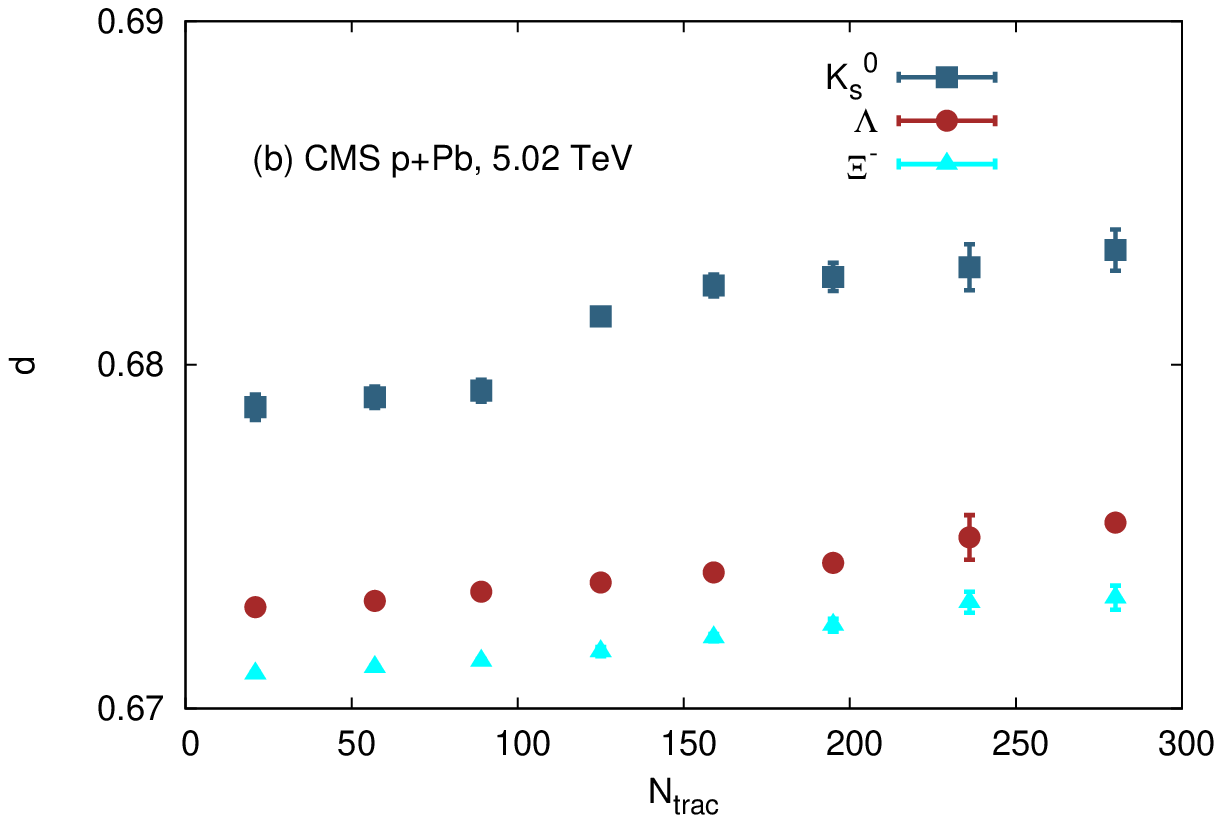}
\includegraphics[scale=0.4]{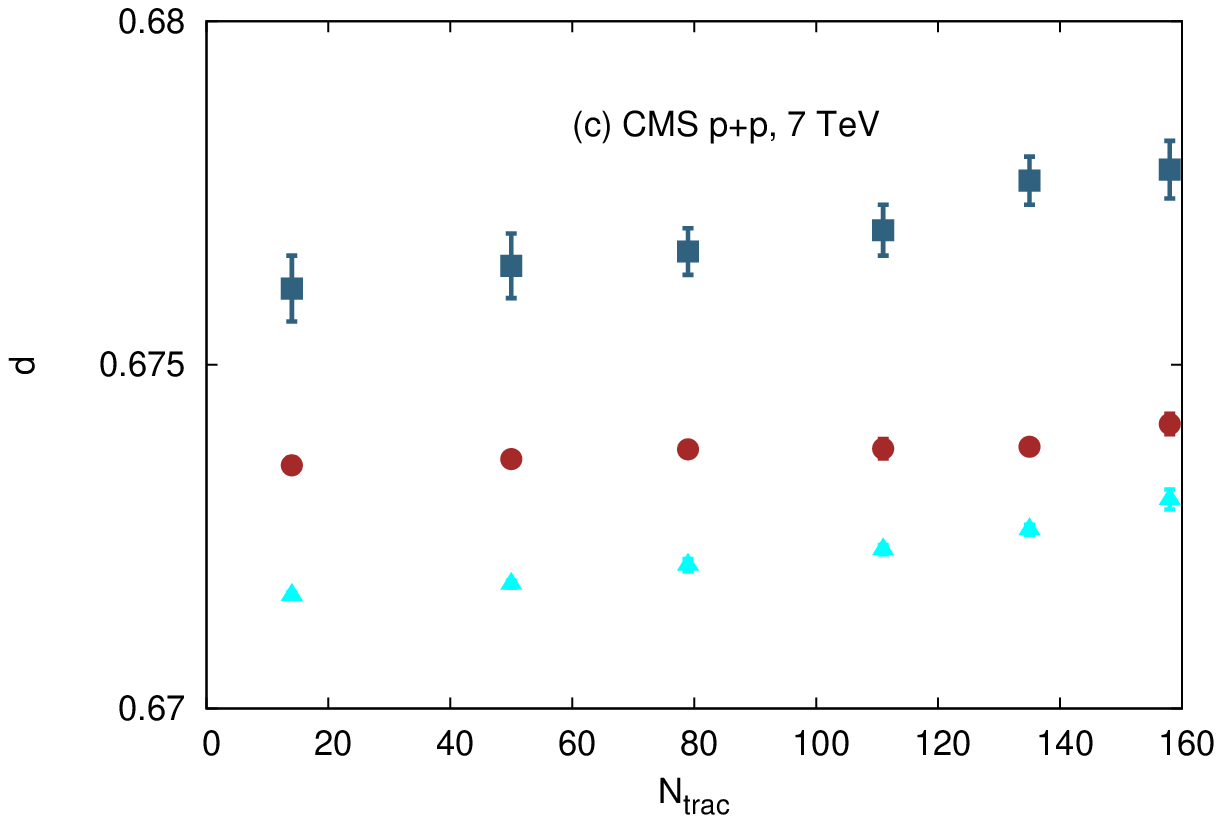}
\caption{(Color online) The (non)extensivity parameter $d$ as deduced from the statistical fits for the strange particles \Kslxi in dependence on $\langle N_{\mathrm{track}}\rangle$ in (a) $\textsf{Pb+Pb}$, at $\sqrt{s_{\mathrm{NN}}}=2.76~$TeV, in (b) $\textsf{p+Pb}$, at $\sqrt{s_{\mathrm{NN}}}=5.02~$TeV, and in (c) $\textsf{p+p}$ collisions, at $\sqrt{s_{\mathrm{NN}}}= 7~$TeV. The symbols refers to the statistical approaches as in the previous figures. \label{parameters_Ntrack_d} }
\end{center}
\end{figure}

In Figs. \ref{fit:2.76}, \ref{fit:5.02}, and \ref{fit:7}, we focus on the smallest $p_T$, where both types of statistical approaches work well. There is another overall remark to be mentioned now that the resulting fit parameter $T$ greatly differs from one type of statistical approach to another. The resulting fit parameters as deduced from generic (non)extensive statistical fits for the transverse momentum distributions in different multiplicity intervals from $\textsf{Pb+Pb}$, $\textsf{p+Pb}$ and $\textsf{p+p}$ are depicted in Figs. \ref{parameters_Ntrack_PbPb}-\ref{parameters_Ntrack_d}. The qualities of the statistical fits, at $c=0.99988$, $\mu=0$, and $\sqrt{s_{\mathtt{\mathrm{NN}}}}=2.76$, $5.02$, $7~$TeV, are summarized in Tab. \ref{Tab1}.

\begin{table}
\vspace{0mm}
\begin{center}
\begin{tabular}{|c|c|c|c|c|c|c|c|c|c|c}
\cline{1-9}
&  \multicolumn{8}{ c|}{$<N_{\mathrm{track}}>$} \\ \cline{2-9}
\multicolumn{1}{ |c| }{\multirow{1}{*}{$\sqrt{s_{\mathtt{\mathrm{NN}}}}$ [TeV]} } &
 $299$ & $253$ & $210$ & $168$ & $130$ & $92$ & $58$ & $21$ & \\ \hline \hline
\multicolumn{1}{ |c| }{\multirow{3}{*}{2.76} } &
\multicolumn{1}{ |c| } {$0.6858$} & $0.8577$ & $0.5994$ & $1.256$ & $1.206$ & $1.174$ & $1.651$ & $2.387$ & $K_s^0$ \\ \cline{2-10}
\multicolumn{1}{ |c  }{}                        &
\multicolumn{1}{ |c| }{$0.9626$} & $1.689$ & $1.676$ & $0.7011$ & $1.111$ & $2.379$ & $2.141$ &$3.364$ & $\Lambda$ \\ \cline{2-10}
\multicolumn{1}{ |c  }{}                        &
\multicolumn{1}{ |c| }{$3.517$} & $2.4$ & $2.47$ & $2.385$ & $3.409$ & $3.186$ & $2.982$ & $2.197$ & $\Xi^-$ \\ \hline \hline 
&  $280$ & $236$ & $195$ & $159$ & $125$ & $89$ & $57$ & $21$ & \\  \cline{2-10} 
\multicolumn{1}{ |c|  }{\multirow{3}{*}{5.02} } &
\multicolumn{1}{ |c| } {$2.13$} & $2.467$ & $1.616$ & $1.33$ & $1.002$ & $1.68$ & $1.797$ & $2.763$ & $K_s^0$ \\ \cline{2-10}
\multicolumn{1}{ |c  }{}                        &
\multicolumn{1}{ |c| }{$2.182$} & $2.188$ & $2.507$ & $1.231$ & $1.353$ & $0.2748$ & $1.119$ & $1.767$ & $\Lambda$ \\ \cline{2-10}
\multicolumn{1}{ |c  }{}                        &
\multicolumn{1}{ |c| }{$4.987$} & $5.061$ & $3.644$ & $2.234$ & $3.392$ & $1.753$ & $1.712$ & $2.355$ & $\Xi^-$ \\ \hline \hline 
&  $158$ & $135$ & $111$ & $79$ & $50$ & $14$ &  & & \\ \cline{2-10}
\multicolumn{1}{ |c|  }{\multirow{3}{*}{7} } &
\multicolumn{1}{ |c| } {$2.301$} & $1.835$ & $2.325$ & $2.289$ & $2.853$ & $4.548$ & & & $K_s^0$ \\ \cline{2-10}
\multicolumn{1}{ |c  }{}                        &
\multicolumn{1}{ |c| }{$2.004$} & $1.267$ & $2.293$ & $1.917$ &  $1.865$ & $3.125$ & & & $\Lambda$ \\ \cline{2-10}
\multicolumn{1}{ |c  }{}                        &
\multicolumn{1}{ |c| }{$2.899$} & $1.546$ & $1.643$  & $2.888$ & $1.758$ & $1.96$ & & & $\Xi^-$ \\ \cline{1-10}
\end{tabular}
\vspace{0mm}
\caption{For $c=0.99988$ and $\mu=0$, the qualities of the generic (non)extensive statistical fits for the transverse momentum distributions ($\chi^2$) are determined for different multiplicity intervals in $\textsf{Pb+Pb}$, $\textsf{p+Pb}$ and $\textsf{p+p}$ collisions, at $\sqrt{s_{\mathtt{\mathrm{NN}}}}=2.76$, $5.02$, $7~$TeV. }
\label{Tab1}
\end{center}
\vspace{0mm}
\end{table}

The present study is designed to determine the possible variations between the three types of statistical approaches and to characterize their dependence on energy, multiplicity, and size of the collisions. Using generic (non)extensive statistical approach, we extract various fit parameters and compare these to the corresponding results obtained from fits to Tsallis and Boltzmann statistics \cite{Yassin:2018svv}. 

For the sake of simplicity, we assumed that the geometry of the fireball is spherical so that the volume reads $V=4/3\pi R^3$, where $R$ defines the dimension of the interacting system. On the other hand, the volume can be related to the normalization of the statistical distribution function, Eq. (\ref{eq:epsln}), which is utilized in describing the particle spectra \cite{Cleymans:2013rfq} or and yields \cite{Tawfik:2014eba}, section \ref{sec:app}.

Figure~\ref{parameters_Ntrack_PbPb} shows the fit parameters $T$ (left) and $V$ (right) as obtained from the strange hadrons \Kslxi as functions of $\langle N_{\mathrm{track}}\rangle$ in $\textsf{Pb+Pb}$ collisions, at $\sqrt{s_{\mathtt{\mathrm{NN}}}}=2.76~$TeV (circles). The results obtained from the generic (non)extensixe statistical approach (cirlcles) are compared with the results deduced from Tsallis (trianlges) and Boltzmann statistics (squares), \cite{Yassin:2018svv}. Fig.~\ref{parameters_Ntrack_pPb} presents the same as in Fig. \ref{parameters_Ntrack_PbPb}, but here $\textsf{p+Pb}$ collisions, at $\sqrt{s_{\mathtt{\mathrm{NN}}}}=5.02~$TeV. Similarly, Fig. \ref{parameters_Ntrack_pp} depicts results from $\textsf{p+p}$ collisions, at $\sqrt{s_{\mathtt{\mathrm{NN}}}}=7~$TeV.

We notice that increasing $\langle N_{\mathrm{track}}\rangle$ is accompanied with an overall raise in both  temperature $T$ and volume $V$. The slope of such an almost linear dependence seems being effected by the type of the statistical approach. Furthermore, we find that the particle mass and its content of strange quantum numbers suppresses the increase in both $T$ and $V$.

The comparison with corresponding results deduced from Boltzmann and Tsallis statistics is also depicted, from which a few remarks are to be highlighted. First, $T_{\mathrm{G}}$ obtained from the generic (non)extensive statistics agrees well with refs.  \cite{Castorina:2014cia,Tawfik:2016tfe} but apparently disagrees with ref. \cite{Yassin:2018svv}. This would mean that the produced particles with large masses and large strange quantum numbers likely freeze out earlier than the ones with smaller masses and less strange quantum numbers \cite{Khuntia:2017ite}. Second, $T_{\mathrm{B}}$ deduced from Boltzmann statistics is greater than the one obtained from Tsallis statistics, $T_{\mathrm{Ts}}$, which in turn is larger than the temperature gained from the generic (non)extensive statistics, $T_{\mathrm{G}}$. This conclusion that $T_{\mathrm{B}} > T_{\mathrm{Ts}} > T_{\mathrm{G}}$ isn't depending on the type of particle or collision. Third, this result would be understood due to the different types of statistics. They apparently manifest transitions from chemical (larger temperature) to the kinetic freezeouts (lower temperature).

The  volume extracted from generic (non)extensive statistics and the ones determined from Tsallis and Boltzmann statistics \cite{Yassin:2018svv} increase with the increase in $\langle N_{\mathrm{track}}\rangle$ for all systems. The volume values deduced from Tsallis and Boltzmann are close to each other. Both are  greater than the volume obtained from generic (non)extensive statistics.  Furthermore, we conclude that the volume in $\textsf{Pb+Pb}$ collisions is greater than the one in $\textsf{p+Pb}$ collisions, which in turn is larger than in $\textsf{p+p}$ collisions. Both conclusions aren't depending on the type of the produced particle.

Figure~\ref{parameters_Ntrack_d} presents the universality (equivalent) class $d$ as deduced for the statistical fits of the strange hadrons \Kslxi  in dependence on $\langle N_{\mathrm{track}}\rangle$ in different types of collisions; (a) $\textsf{Pb+Pb}$, at $\sqrt{s_{\mathtt{\mathrm{NN}}}}=2.76~$TeV, (b) $\textsf{p+Pb}$, at $\sqrt{s_{\mathtt{\mathrm{NN}}}}=5.02~$TeV, and (c) $\textsf{p+p}$ collisions, at $\sqrt{s_{\mathtt{\mathrm{NN}}}}=7~$TeV. We notice that the value of $d$ slightly increases with increasing $\langle N_{\mathrm{track}}\rangle$ in all types of collisions. Also, $d$ decreases with the increase in both particle masses and strange quantum numbers. The observation that $d$ decreases with $\langle N_{\mathrm{track}}\rangle$ apparently manifests that the statistical nature of the system approaches extensivity, especially that the second equivalent class $c$ is taken very close to unity, $c=0.99988$, indicating that the Boltzmann statistics becomes the proper statistical approach in describing that system.

In the next section, we study the possible correlations between the resulting fit parameters, especially with the collision energies.

\subsection{Correlations between resulting fit parameters}
\label{res2}

\begin{figure}[h]
\begin{center}
\includegraphics[scale=0.4]{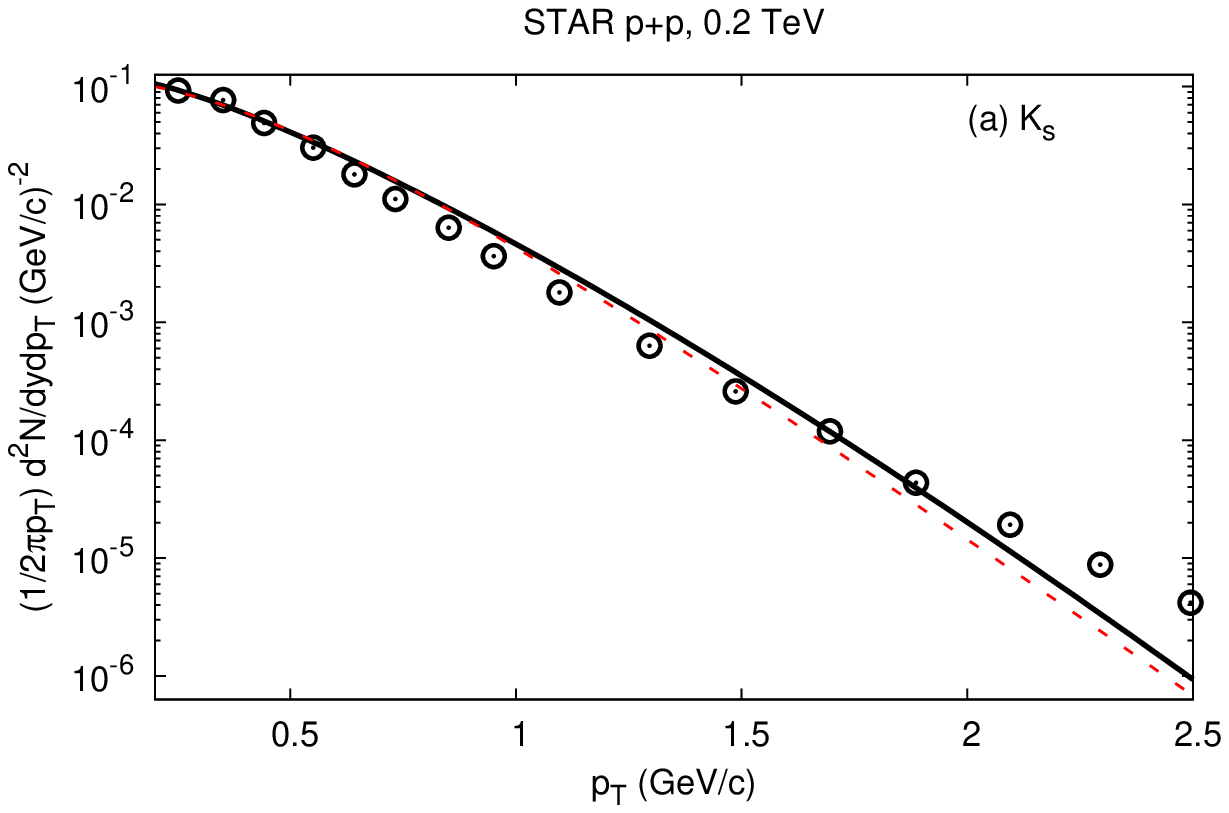}
\includegraphics[scale=0.4]{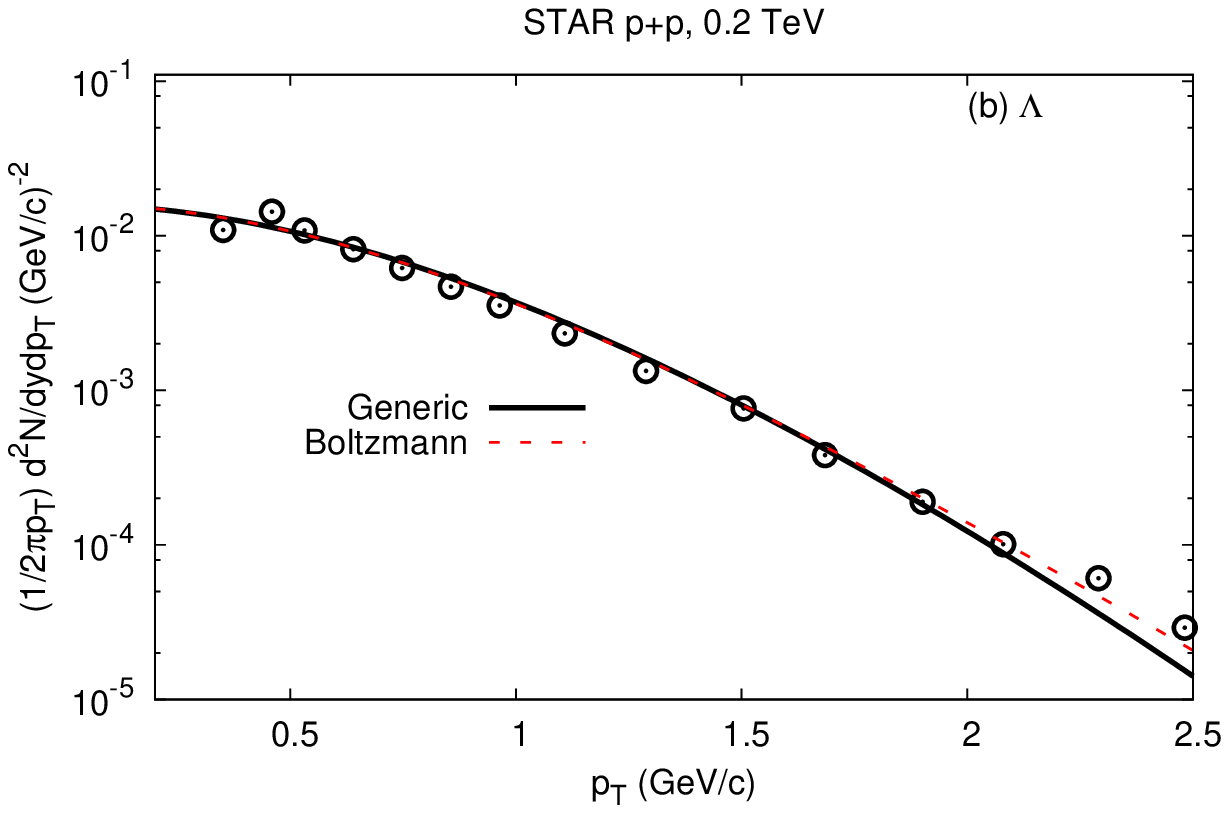}
\includegraphics[scale=0.4]{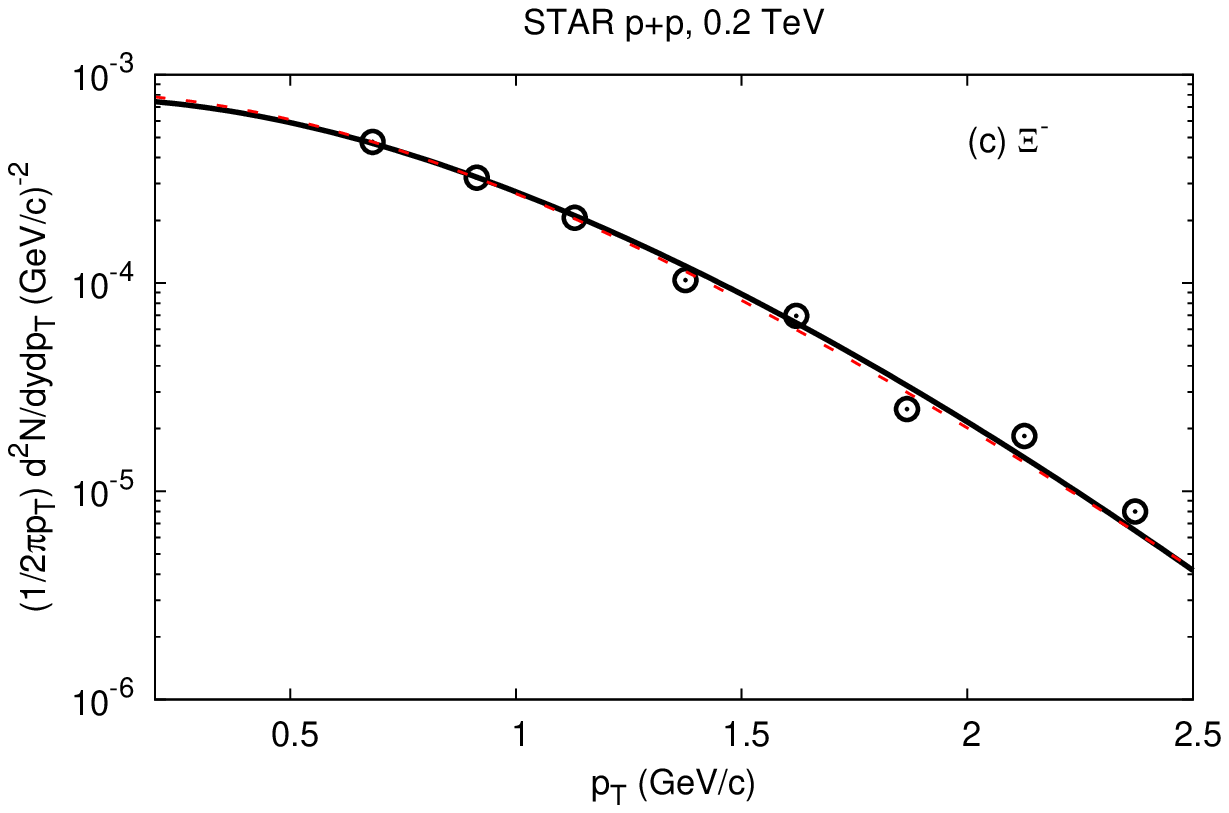}
\caption{(Color online) Transverse momentum distributions of the strange hadrons \Kslxi for $\textsf{p+p}$ collisions, at $\sqrt{s_{\mathrm{NN}}}=0.2~$TeV, measured in most central \textit{STAR} collisions (symbols) \cite{Adam:2019koz} are compared with the calculations based on generic (non)extensive (solid curve), Eq.~\ref{transmomentum1}, and Boltzmann statistics (dashed curve), Eq.~\ref{eq4}. }
\label{fit:200}
\end{center}
\end{figure}

\begin{figure}[h]
\begin{center}
\includegraphics[scale=0.4]{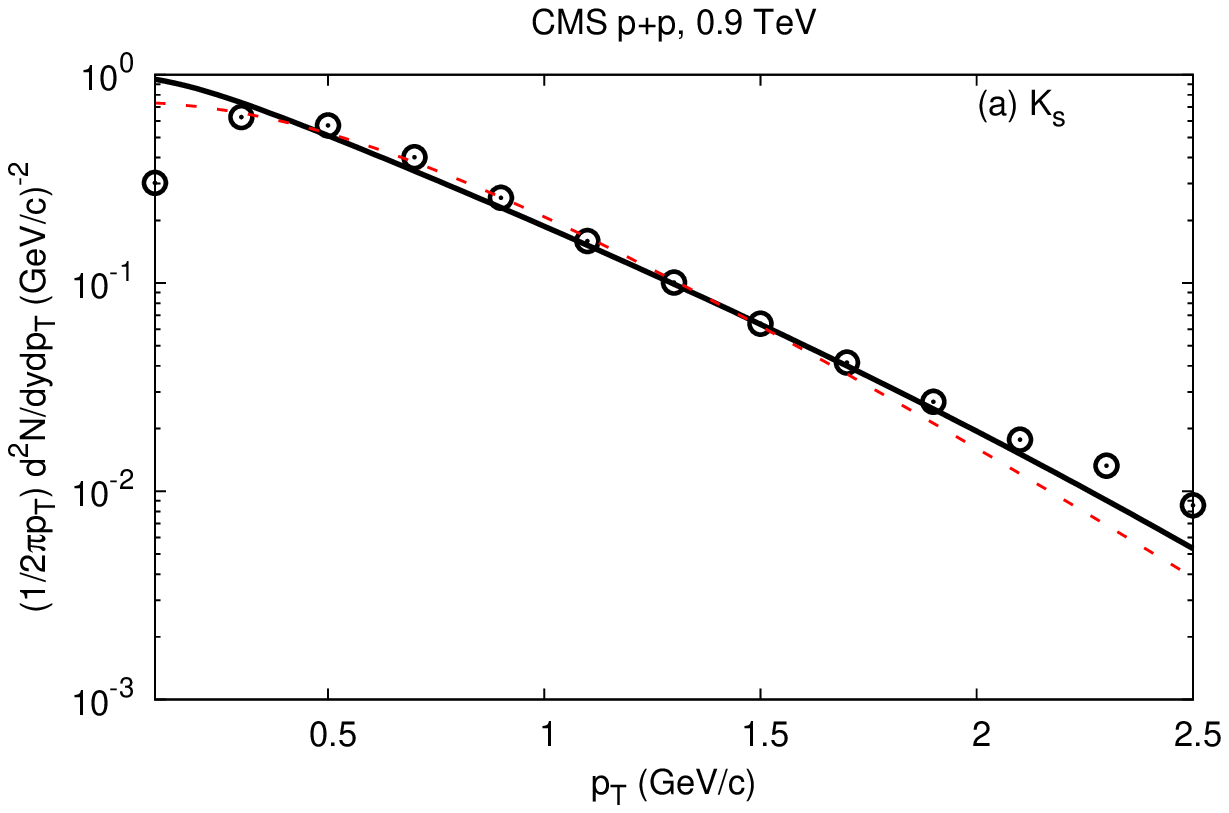}
\includegraphics[scale=0.4]{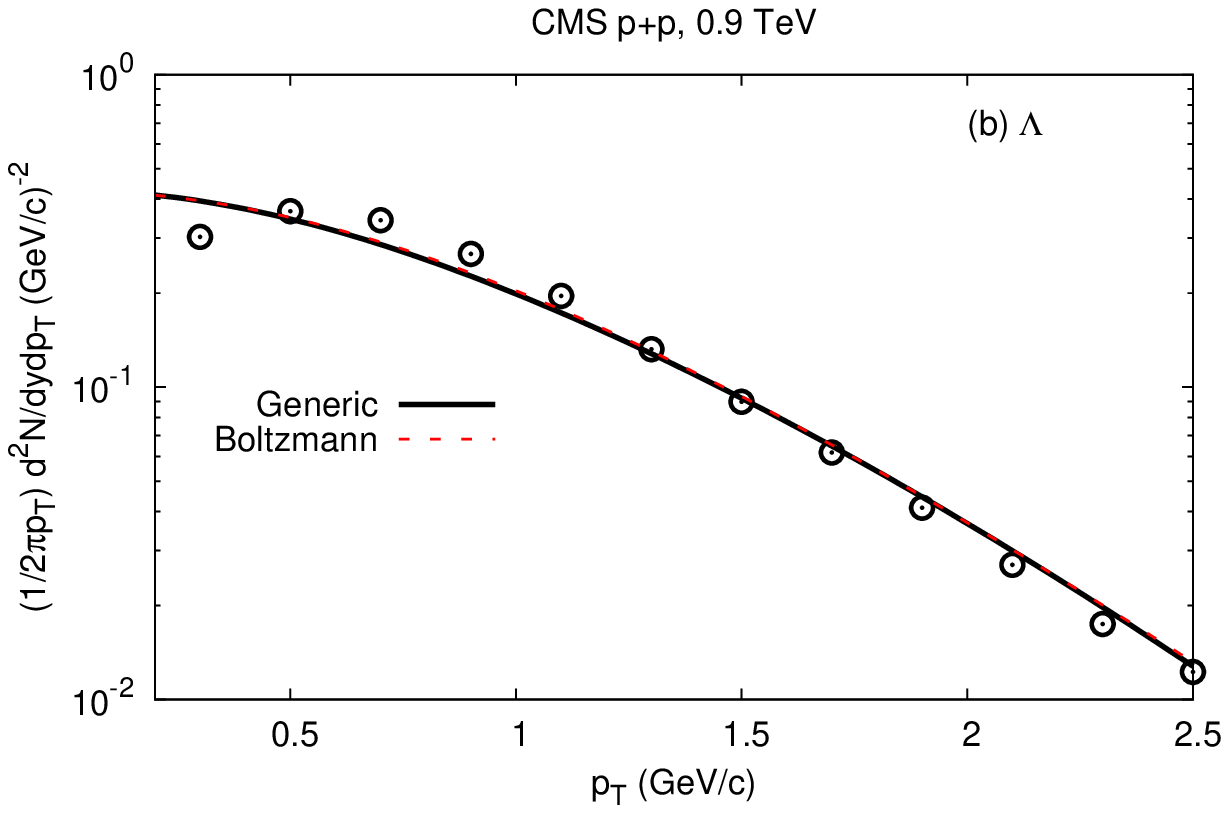}
\includegraphics[scale=0.4]{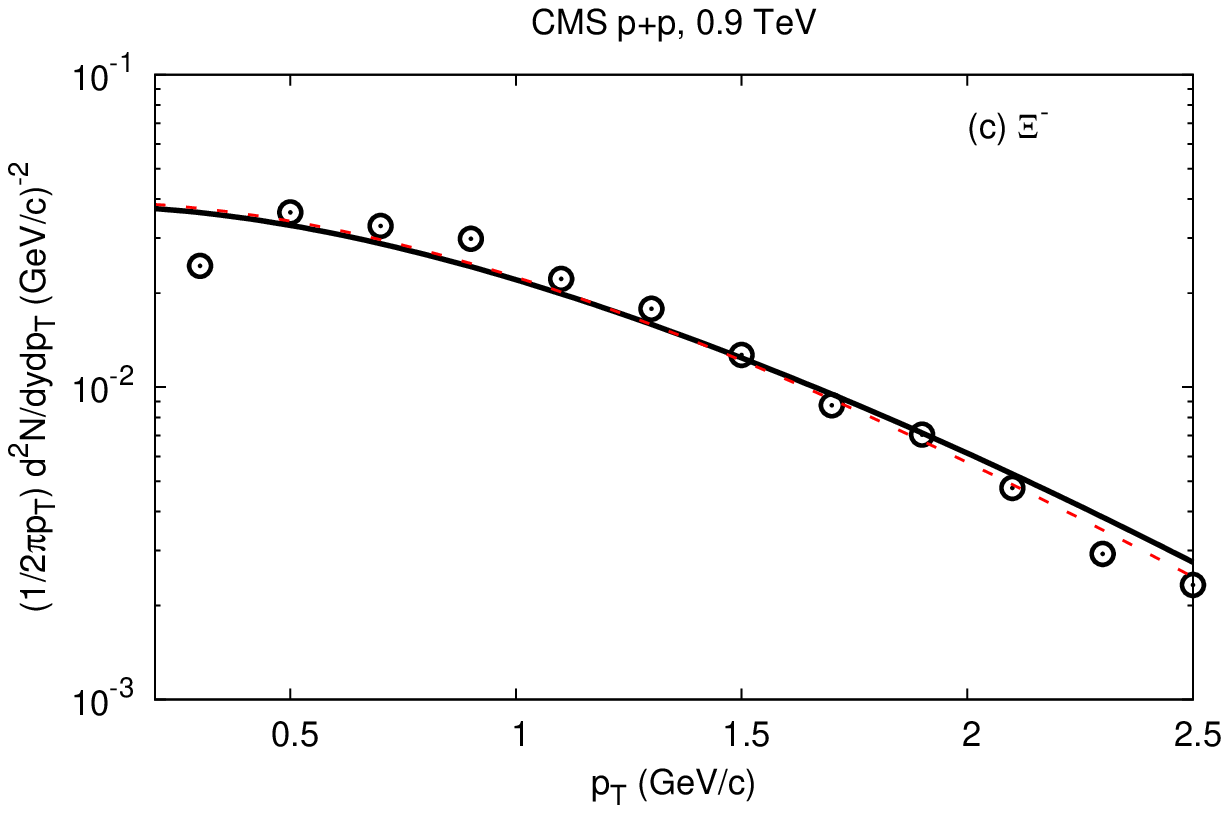}
\caption{(Color online) The same as in Fig. \ref{fit:200}, but here for the most central $(p+p)$ collisions in the \textit{CMS} experiment (symbols) \cite{Khachatryan:2011tm}, at $\sqrt{s_{\mathrm{NN}}}=0.9~$TeV. The generic (non)extensive, Eq.~\ref{transmomentum}, and Boltzmann statistics, Eq.~\ref{eq3}, are illustrated by solid and dashed curves, respectively. \label{fit:900} }
\end{center}
\end{figure}

\begin{figure}[t!]
\begin{center}
\includegraphics[scale=0.4]{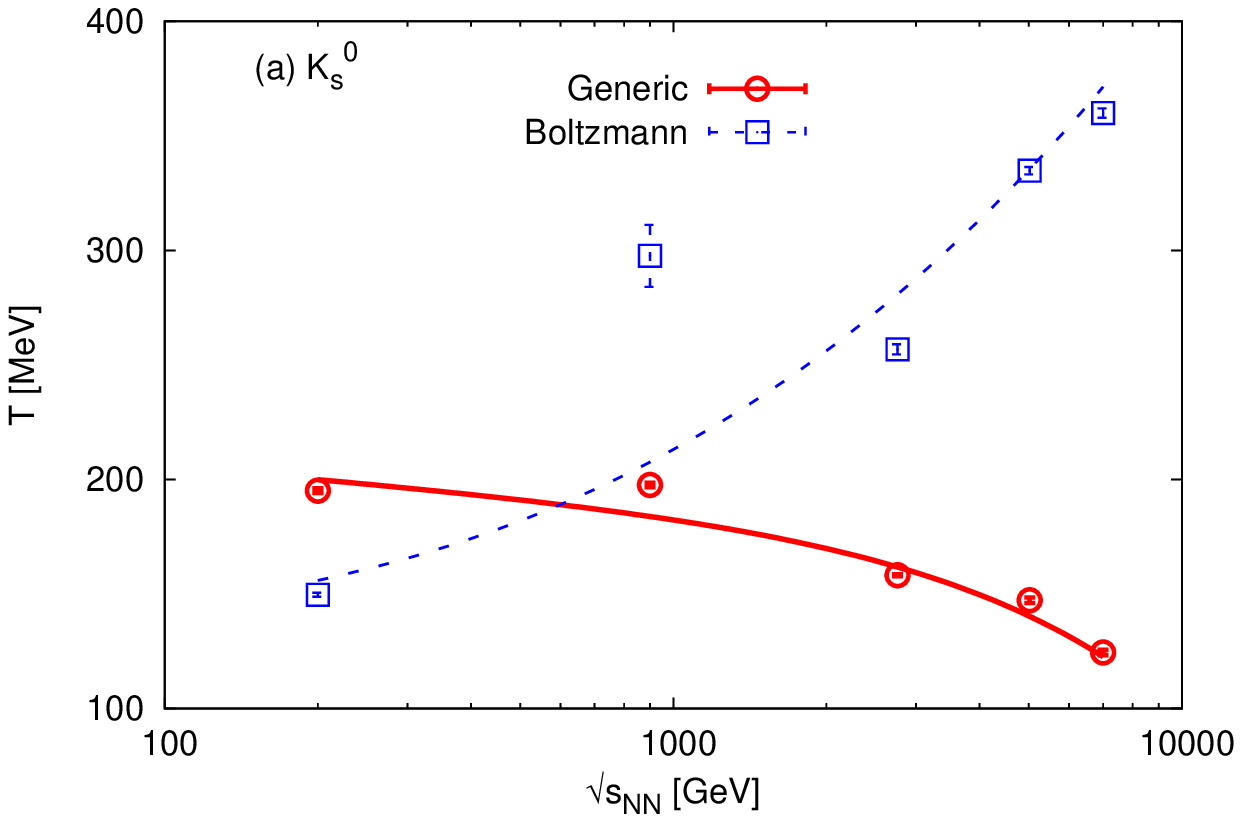}
\includegraphics[scale=0.4]{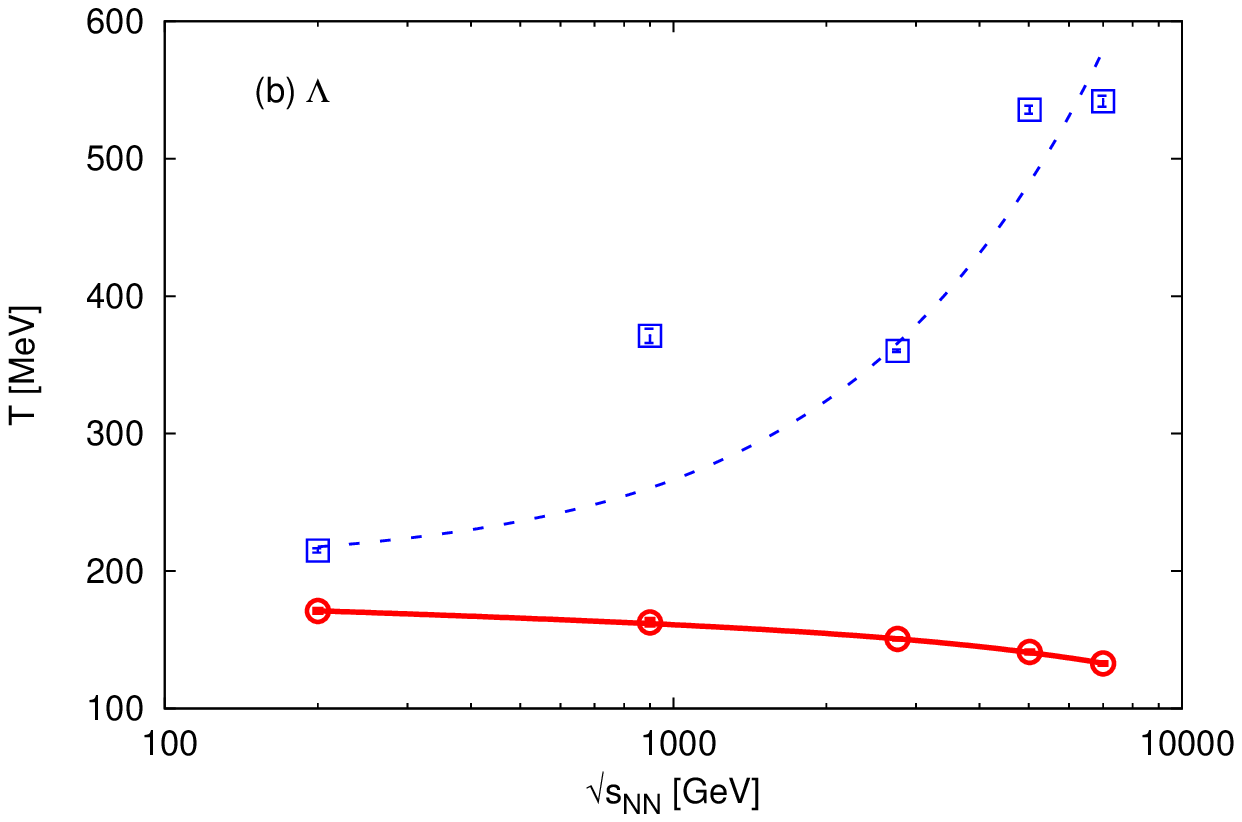}
\includegraphics[scale=0.4]{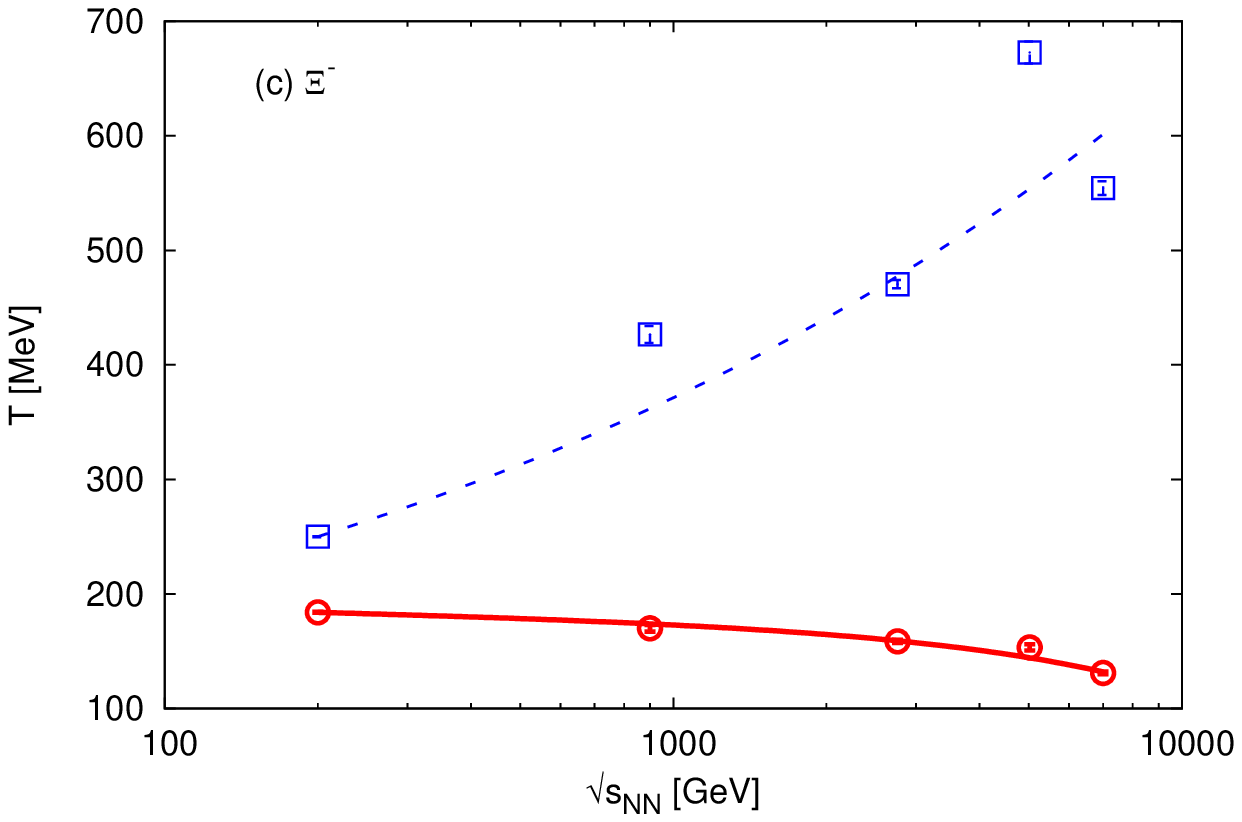}
\caption{(Color online) The fit parameter $T$ as functions of $\sqrt{s_{\mathtt{\mathrm{NN}}}}$ for the strange hadrons (a) $K_s^0$, (b) $\Lambda$, and (c) $\Xi^-$ fitted to Boltzmann and generic (non)extensive statistics. The temperature values are obtained from the statistical analysis of $p_T$ in most central collisions, at different energies, as in Figs. \ref{fit:2.76}-\ref{fit:7}, \ref{fit:200} and \ref{fit:900}.
\label{T_sqrt} }
\end{center}
\end{figure}

\begin{figure}[t!]
\begin{center}
\includegraphics[scale=0.4]{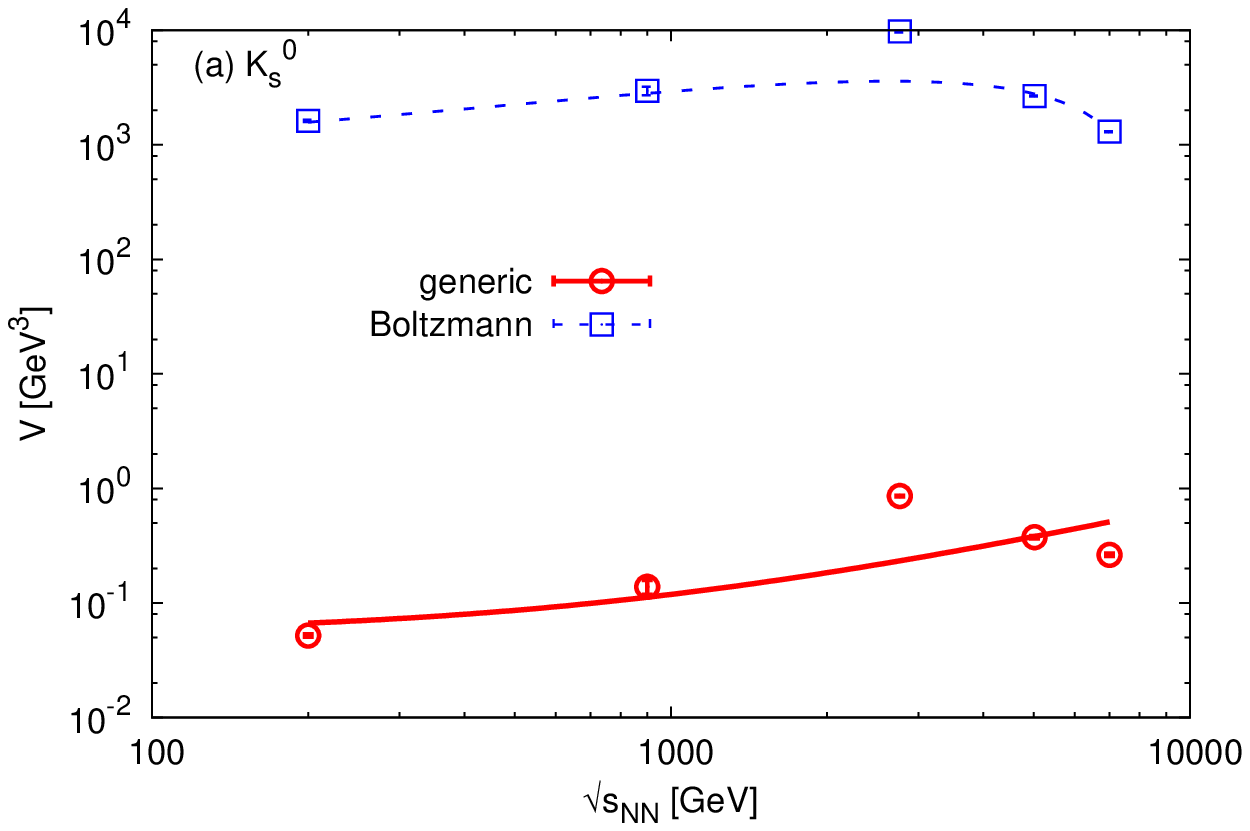}
\includegraphics[scale=0.4]{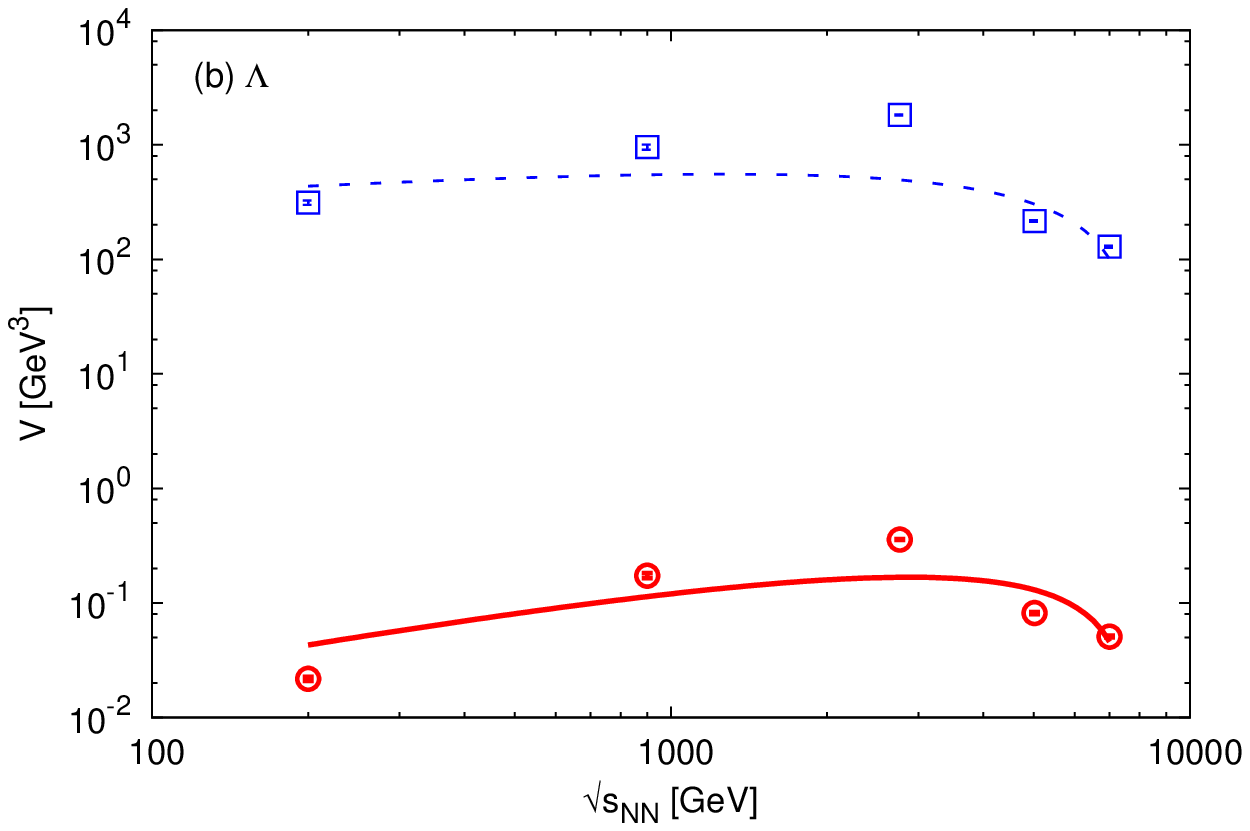}
\includegraphics[scale=0.4]{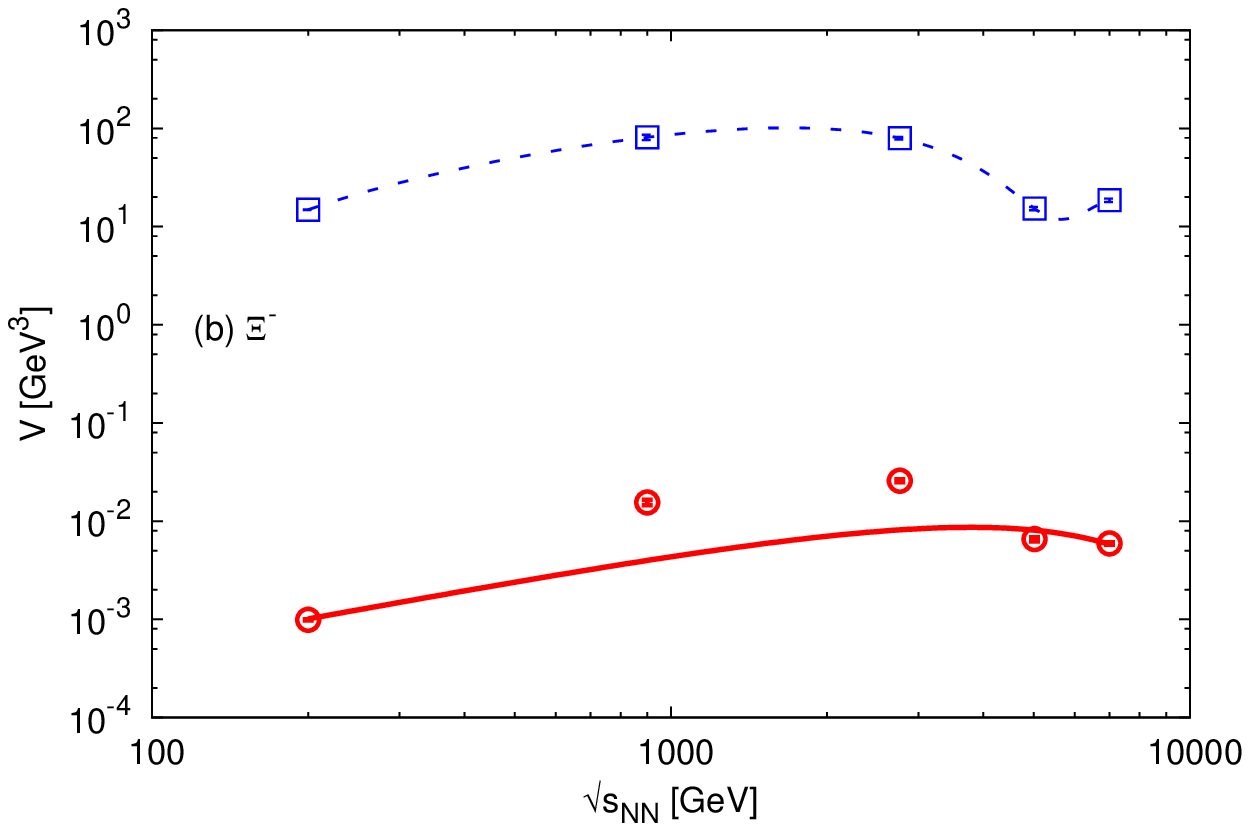}
\caption{(Color online) The same as Fig. \ref{T_sqrt}, but here for the volume $V$  obtained from the statistical analysis of $p_T$ in most-central collisions, at different energies, as in Figs. \ref{fit:2.76}-\ref{fit:7}, \ref{fit:200} and \ref{fit:900}.  \label{V_sqrt} }
\end{center}
\end{figure}

\begin{figure}[t!]
\begin{center}
\includegraphics[scale=0.75]{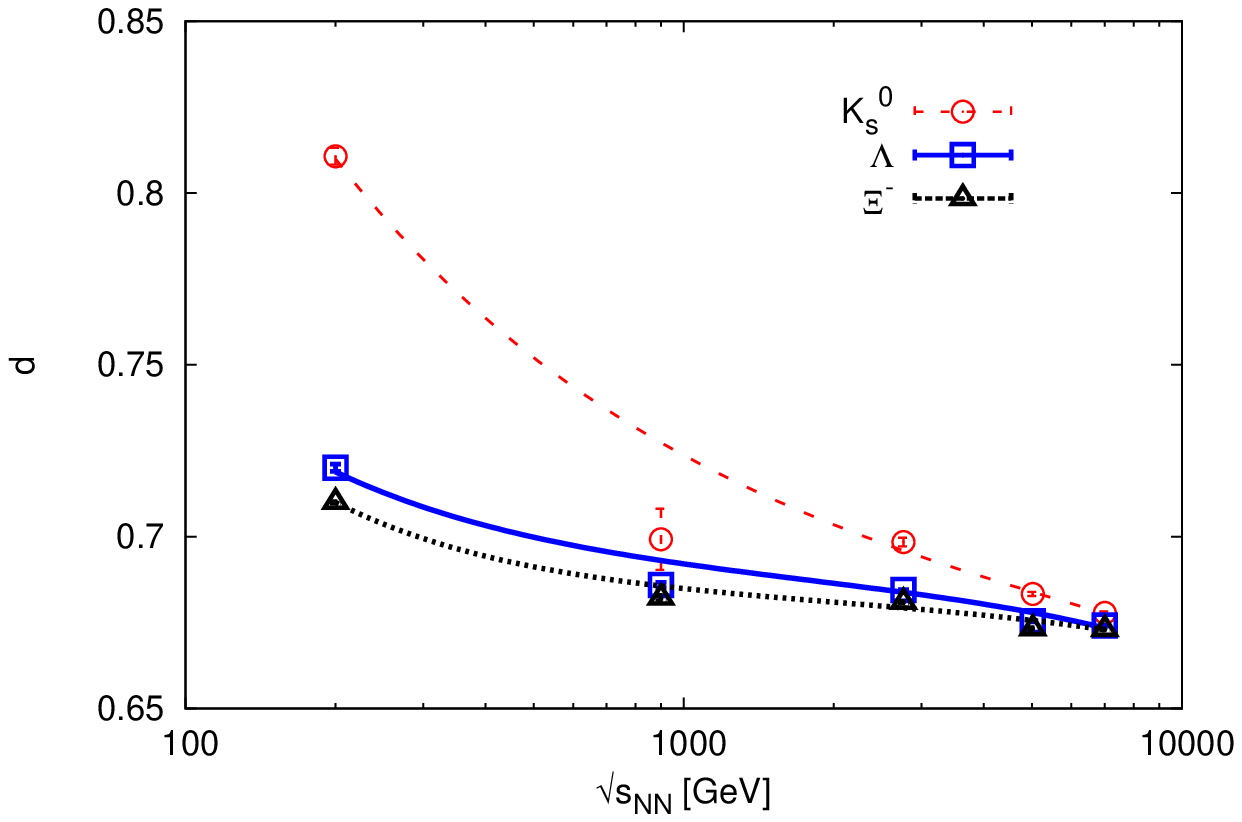}
\caption{(Color online) The nonextensive parameter $d$ in dependence on $\sqrt{s_{\mathtt{\mathrm{NN}}}}$ as estimated from the statistical fits for the strange hadrons \Kslxi to the generic (non)extensive statistics.The values of $d$ are obtained from the statistical analysis of $p_T$ in the most-central collisions, at different energies, as in Figs. \ref{fit:2.76}-\ref{fit:7}, \ref{fit:200} and \ref{fit:900}. \label{d_sqrt} }
\end{center}
\end{figure}

In this section, we propose expressions relating the dependence of the resulting fit parameters on the collision energies. To this end, we analyze, for instance, the transverse momentum distribution for the same strange hadrons measured in most central collisions, at different energies.  Fig.~\ref{fit:200} presents the transverse momentum $p_{\mathrm{T}}$ spectra of, (a) $K_s^0$, (b) $\Lambda$, and (c) $\Xi^-$, produced in most central $\textsf{p+p}$ collisions in the mid-rapidity range $\vert y \vert<1.0$, at $\sqrt{s_{\mathrm{NN}}}=0.2~$TeV. The symbols refer to the experimental results \cite{Adam:2019koz}. Our calculations using generic (non)extensive and Boltzmann statistics are depicted as solid and dashed curves, respectively. Here, we focus on the smallest $p_{\mathrm{T}}$ region, where $c=0.99988$, and $\mu \approx 25~$MeV. We find that our calculations for $\Lambda$ and $\Xi^-$ agree well with the experimental $\textsf{p+p}$ results. Fig.~\ref{fit:900} shows the same as Fig. \ref{fit:200}, but here at $\sqrt{s_{\mathrm{NN}}}=0.9~$TeV and  $\vert y \vert<1.0$. The symbols represent the \textit{CMS} results \cite{Khachatryan:2011tm}. We also find that our calculations for $\Lambda$ and $\Xi^-$ agree well with $\textsf{p+p}$ collisions, at $c=0.99988$ and $\mu=0$. Our calculations using both extensive and generic (non)extensive statistics seem to agree well with the experimental data. We can extract various fit parameters from both statistics, at a wide range of energies, as shown in Figs. \ref{T_sqrt}-\ref{d_sqrt}.

Figure \ref{T_sqrt} depicts the dependence of the fit parameter $T$ on the collision energies $\sqrt{s_{\mathtt{\mathrm{NN}}}}$ for the strange hadrons \Kslxi as obtained from generic (non)extensive and Boltzmann statistics. For all studied strange hadrons, we find that the temperature determined from generic (non)extensive fits decreases with the increase in $\sqrt{s_{\mathtt{\mathrm{NN}}}}$, i.e. increasing collision centrality, while the temperarue obtained from the Boltzmann fits increases with the increase in $\sqrt{s_{\mathtt{\mathrm{NN}}}}$. Also, we find that the generic (non)extensive temperature decreases with the increase in both particle masses and strange qunatum numbers, while the Boltzmann temperature shows an opposite dependence. On the other hand, it is worthy highlighting that the generic (non)extensive temperature obtained agrees well with previous studies \cite{Castorina:2014cia,Tawfik:2016tfe}, in which it was concluded that the freezeout of such particles with large masses and large strange quantum numbers takes place earlier than the one for light particles with small strange quantum numbers. \cite{Khuntia:2017ite}.  The Boltzmann temperature and volume are deduced from the statistical analysis of $p_T$ spectra, at $200$ and $900~$GeV. The dependence of temperature on the energy seems varying with the type of statistics but being independent on the system size of the collision. To summarize we find that $T_B$ is to be related to the {\it chemical} freezeout temperature, while $T_G$ to the {\it kinetic} freezeout temperature. This is based on phenomenological point-of-view, where the temperature obtained from generic (non)extensive statistics can be related to the kinetic freezeout temperature, while the temperature obtained from Boltzmann extensive statistics looks similar to the chemical freezeout temperature.

Figure \ref{V_sqrt} presents the fit parameter $V$ as a function of $\sqrt{s_{\mathtt{\mathrm{NN}}}}$ for the strange hadrons \Kslxi as fitted to generic (non)extensive and Boltzmann statistics. For all hadrons, we find that the volume deduced from both types of statistics has almost the same behavior. On the other hand, the Boltzmann volume is found greater than the generic (non)extensive volume. Also, it is apparent that the volume deduced from both types of statistics increases with the increase in $\sqrt{s_{\mathtt{\mathrm{NN}}}}$ till $2.76~$GeV, the volume decreases with the increase in $\sqrt{s_{\mathtt{\mathrm{NN}}}}$. This trend indicates that the dependence on the energy is only more effective, at low energy. It is noted that the value of the volume from both statistics decreases with the increase in the particle masses and strange quantum numbers which lead to the dependence of the number of large particles (mass and strange content). In other words, the volume of the system is small if this system contains hadrons with large mass and/or large strange quantum numbers.

Figure \ref{d_sqrt} shows the nonextensive parameter $d$ as a function of  $\sqrt{s_{\mathtt{\mathrm{NN}}}}$ for the strange hadrons \Kslxi, which have been fitted to generic (non)extensive statistics. For all strange hadrons, we find that $d$ decreases with the increase in $\sqrt{s_{\mathtt{\mathrm{NN}}}}$. Also, we observe that $d$ decreases with the increase in both particle masses and strange quantum numbers. The decrease in $d$ with $\sqrt{s_{\mathtt{\mathrm{NN}}}}$ indicates that the system starts from a non-equilibrium state, nonextensive, at low $\sqrt{s_{\mathtt{\mathrm{NN}}}}$, and apparently approaches an equilibrium state, extensive, at large $\sqrt{s_{\mathtt{\mathrm{NN}}}}$, i.e. a novel kind of a statistical transition, which can only be characterized by generic (non)extensive statistics.

\subsubsection{Analytical expressions for the resulting fit parameters}
\label{sec:outExprs}

Tables \ref{tab:fit_Boltzmann} and \ref{tab:fit_Generic} and figures \ref{T_sqrt}, \ref{V_sqrt}, and \ref{d_sqrt} list out and depict various fit parameters. In the following, we summarize the various dependences of these thermodynamical quantities on the collision energy.
\begin{itemize}
\item Using Boltzmann statistics, the dependence of temperature on $\sqrt{s_{\mathtt{\mathrm{NN}}}}$ for all particles is given as
\begin{equation}
T = (a (\sqrt{s_{\mathtt{\mathrm{NN}}}} + b))^c,
\label{proexp1}
\end{equation}
where the values of $a$, $b$, and $c$ are taken from Fig. \ref{T_sqrt}, see Tab. \ref{tab:fit_Boltzmann}.
\item The dependence of volume on $\sqrt{s_{\mathtt{\mathrm{NN}}}}$ for $K_s^0$ can be expressed as
\begin{equation}
V = (a (\sqrt{s_{\mathtt{\mathrm{NN}}}}+b))^c + f \sqrt{s_{\mathtt{\mathrm{NN}}}},
\label{proexp2}
\end{equation}
where the values of $a$, $b$, $c$, and $f$ are taken from Fig. \ref{V_sqrt}, see Tab. \ref{tab:fit_Boltzmann}.
\item For $\Lambda$:
\begin{equation}
V = a \sqrt{s_{\mathtt{\mathrm{NN}}}}^b + c \sqrt{s_{\mathtt{\mathrm{NN}}}},
\label{proexp3}
\end{equation}
where the values of $a$, $b$, and $c$ are taken from Fig. \ref{V_sqrt}, see Tab. \ref{tab:fit_Boltzmann}.
\item For $\Xi^-$:
\begin{equation}
V = a \sqrt{s_{\mathtt{\mathrm{NN}}}} + b \sqrt{s_{\mathtt{\mathrm{NN}}}}^2 + c \sqrt{s_{\mathtt{\mathrm{NN}}}}^3 + f \sqrt{s_{\mathtt{\mathrm{NN}}}}^4 + g,
\label{proexp4}
\end{equation}
where the values of $a$, $b$, $c$, $f$, and $g$ are taken from Fig. \ref{V_sqrt}, see Tab. \ref{tab:fit_Boltzmann}.

\item By using the generic (non)extensive statistical approach, we have estimated the dependence of  temperature on $\sqrt{s_{\mathtt{\mathrm{NN}}}}$ for all particles as
\begin{equation}
T = a \sqrt{s_{\mathtt{\mathrm{NN}}}}^b  + c \sqrt{s_{\mathtt{\mathrm{NN}}}},
\label{proexp5}
\end{equation}
where the values of $a$, $b$, and $c$ are shown in Fig. \ref{T_sqrt}, see Tab. \ref{tab:fit_Generic}.
\item For $K_s^0$, the dependence of volume on $\sqrt{s_{\mathtt{\mathrm{NN}}}}$ is expressed as
\begin{equation}
V = (a (\sqrt{s_{\mathtt{\mathrm{NN}}}} + b))^c + f \sqrt{s_{\mathtt{\mathrm{NN}}}},
\label{proexp6}
\end{equation}
where the values of $a$, $b$, $c$, and $f$ are taken from Fig. \ref{V_sqrt}, see Tab. \ref{tab:fit_Generic}.
\item For $\Lambda$:
\begin{equation}
V = a \sqrt{s_{\mathtt{\mathrm{NN}}}}^b  + c \sqrt{s_{\mathtt{\mathrm{NN}}}},
\label{proexp7}
\end{equation}
where the values of $a$, $b$, and $c$ are taken from Fig. \ref{V_sqrt}, see Tab. \ref{tab:fit_Generic}.
\item For $\Xi^-$:
\begin{equation}
V = a \sqrt{s_{\mathtt{\mathrm{NN}}}} + b \sqrt{s_{\mathtt{\mathrm{NN}}}}^2 + c \sqrt{s_{\mathtt{\mathrm{NN}}}}^3,
\label{proexp8}
\end{equation}
where the values of $a$, $b$, and $c$ are taken from Fig. \ref{V_sqrt}, see Tab. \ref{tab:fit_Generic}.

\item For all particles, the dependence of the equivalent class $d$ on $\sqrt{s_{\mathtt{\mathrm{NN}}}}$ is suggested as
\begin{equation}
d = (a + \sqrt{s_{\mathtt{\mathrm{NN}}}}^{-b})^c + f \sqrt{s_{\mathtt{\mathrm{NN}}}},
\label{proexp9}
\end{equation}
where the values of $a$, $b$, $c$, and $f$, are shown in Fig. \ref{d_sqrt}, see Tab. \ref{tab:fit_Generic}.
\end{itemize}
So far, we conclude that the fit parameters obtained depend on the collision energy and on the type of the statistical approach applied, especially when moving from extensive to nonextensive statistical approach.

\begin{table}[h!]
\centering
  \begin{tabular}{|c|c|c|c|c|c|}  \hline
\multicolumn{1}{ |c|  }{\multirow{1}{*}{Fit Parameters}} &
   & $K_s^0$ & $\Lambda$ & $\Xi^-$  \\ \hline
    \multirow{5}{*}{$T$} & $a$ & $14502.6\pm2916$ &  $0.3873\pm0.0154$ &  $2.242\times10^{7}\pm1.17\times10^{6}$  \\
     \cline{2-5} & $b$ & $285.3022\pm192.8$ &  $2484.34\pm197.6$ &  $4.1853\pm11.08$  \\
      \cline{2-5} & $c$ & $0.3203\pm0.0037$ &  $0.7749\pm0.004$ &  $0.2482\pm0.0006$  \\ \hline
    \multirow{5}{*}{$V$} & $a$ & $51.586\pm0.326$ &  $121.28\pm9.519$ &  $0.166\pm2.762\times10^{-7}$  \\
     \cline{2-5} & $b$ & $61.48\pm33.3$ &  $0.2537\pm0.009$ &  $-7.561\times10^{-5}\pm3.962\times10^{-11}$  \\
      \cline{2-5} & $c$ & $0.834\pm0.0004$ &  $-0.1494\pm0.0139$ &  $1.174\times10^{-8}\pm6.421\times10^{-15}$  \\
       \cline{2-5} & $f$ & $-5.995\pm0.033$ &  $-$ &  $-6.051\times10^{-13}\pm9.83\times10^{-19}$  \\ 
        \cline{2-5} & $g$ & $-$ &  $-$ &  $-15.482\pm0.0004$  \\ \hline
\end{tabular}
\caption{The fit parameters obtained from Boltzmann statistics, Eqs. \ref{proexp1}-\ref{proexp4}. \label{tab:fit_Boltzmann} }
\end{table}

\begin{table}[h!]
\centering
  \begin{tabular}{|c|c|c|c|c|c|}   \hline
\multicolumn{1}{ |c|  }{\multirow{1}{*}{Fit Parameters}} &
   & $K_s^0$ & $\Lambda$ & $\Xi^-$  \\ \hline
    \multirow{5}{*}{$T$} & $a$ & $243.783\pm68.87$ &  $199.109\pm2.364$ &  $208.678\pm16.78$  \\ 
     \cline{2-5} & $b$ & $-0.036\pm0.045$ &  $-0.0279\pm0.0018$ &  $-0.0226\pm0.0152$  \\
      \cline{2-5} & $c$ & $-0.0078\pm0.0046$ &  $-0.003\pm0.0001$ &  $-0.0055\pm0.0015$  \\ \hline
    \multirow{5}{*}{$V$} & $a$ & $2531.7\pm17580$ &  $0.0112\pm2.032$ &  $5.245\times10^{-6}\pm1.179\times10^{-6}$  \\
     \cline{2-5} & $b$ & $1266.2\pm91950$ &  $0.9946\pm1.015$ &  $-9.596\times10^{-10}\pm9.045\times10^{-10}$  \\
      \cline{2-5} & $c$ & $-0.1692\pm0.0679$ &  $-0.0107\pm2.034$ &  $4.702\times10^{-14}\pm1.19\times10^{-13}$  \\
       \cline{2-5} & $f$ & $6.577\times10^{-5}\pm4.065\times10^{-5}$ &  $-$ &  $-$  \\ \hline
          \multirow{5}{*}{$d$} & $a$ & $0.853\pm0.0004$ &  $0.918\pm0.0002$ &  $0.967\pm6.478\times10^{-5}$  \\
     \cline{2-5} & $b$ & $0.504\pm0.003$ &  $0.871\pm0.028$ &  $1.065\pm0.01$  \\
      \cline{2-5} & $c$ & $2.615\pm0.025$ &  $4.408\pm0.026$ &  $11.491\pm0.032$  \\
       \cline{2-5} & $f$ & $-9.113\pm1.348\times10^{-7}$ &  $-2.02\pm0.152\times10^{-6}$ &  $-1.253\pm0.096\times10^{-6}$  \\ \hline
    \end{tabular}
      \caption{The fit parameters obtained from generic (non)extensive statistics, Eqs. \ref{proexp5}-\ref{proexp9}. \label{tab:fit_Generic}  }
\end{table}

\section{Conclusions}
\label{sec:concl}

Within generic (non)extensive statistics, we have analysed the transverse momentum distribution $p_{\mathrm{T}}$ of the strange hadrons \Kslxi  in different multiplicity classes measured  in the $\textit{CMS}$ experiment in $\textsf{Pb+Pb}$ collisions, at $\sqrt{s_{\mathrm{NN}}}=2.76$ TeV, in $\textsf{p+Pb}$ collisions, at $\sqrt{s_{\mathrm{NN}}}=5.02$ TeV, and in $\textsf{p+p}$ collisions, at $\sqrt{s_{\mathrm{NN}}}=0.2$, $0.9$, and $7~$TeV. The fit parameters deduced are compared with previous studies based on Tsallis and Boltzmann statistics \cite{Yassin:2018svv}. This comprehensive comparison indicates variations between the three types of statistical approaches. 

We conclude that the temprature obtained from the generic (non)extensive statistics agrees well with previous studies \cite{Castorina:2014cia,Tawfik:2016tfe} but not with others \cite{Yassin:2018svv}. Thus, we conclude that the produced strange particles with large masses and large strange quantum numbers seem to freeze out earlier than the ones with smaller masses and less strange quantum numbers \cite{Khuntia:2017ite}. As for the dependence on the type of the statistical approaches, we have obtained that the temperature deduced from Boltzmann statistics is greater than the one obtained from Tsallis statistics. The latter is in turn larger than the one determined from generic (non)extensive statistics. This conclusion isn't affected by the type of particles or collisions. As an explanation, we suggest that the different types of statistics play an essential role. The different temperatures obtained would be conjectured to manifest transitions from chemical (larger temperature) to kinetic freezeouts (lower temperature). Accordingly, Boltzmann and Tsallis statistics can be related to chemical and kinetic freezeout, respectively. The generic (non)extensive statistics combines both types of statistical nature. The values assigned to the universality (equivalent) classes $(c,d)$ autonomously dictates statistical nature of the studies system; $(c,d)=(1,1)$ Boltzmann and $(c,d)=(q,0)$ Tsallis, i.e. within the ranges of the classes $(c,d)$ both Boltzmann and Tsallis are very special cases.

As for the volume extracted from generic (non)extensive statistics and compared with Tsallis and Boltzmann statistics, we find that the volume apart from the type of the produced particles increases with the collision centrality for all collision sizes. While the volume values deduced from Tsallis and Boltzmann are close to each other, both are greater than the ones obtained from generic (non)extensive statistics.  We found that the temperature and the volume obtained within the different statistical approaches increase with the increase in the multiplicity classes for all types of collisions. 

As for the energy dependence of the various fit parameters, we first found that the temperature deduced from generic (non)extensive fits decreases with the increase in $\sqrt{s_{\mathtt{\mathrm{NN}}}}$, while the Boltzmann temperature increases.The generic (non)extensive temperature decreases with the particle masses and the strange quantum numbers, while the Boltzmann temperature has an opposite dependence. The earlier refers to early freezeout of particles with large masses and strange quantum numbers. Lighter particles with smaller strange quantum numbers freeze out afterwards. The system size seems not affecting this conclusion. From phenomenological point-of-view, we conclude that the temperature obtained from generic (non)extensive statistics can be compared with the kinetic freezeout temperature, while the Boltzmann temperature can be related to the chemical freezeout temperature.

The nonextensive parameter $d$ decreases with the increase in $\sqrt{s_{\mathtt{\mathrm{NN}}}}$, particle masses, and strange contents indicating that the system moves from a non-equilibrium state, nonextensive, at low $\sqrt{s_{\mathtt{\mathrm{NN}}}}$, towards an equilibrium state, extensive, at large $\sqrt{s_{\mathtt{\mathrm{NN}}}}$. This novel statistical transition can only be characterized by the generic (non)extensive statistics.

\bibliographystyle{aip}
\bibliography{generic_pT1}

\end{document}